\documentclass[aps,prd,twocolumn,showpacs,superscriptaddress]{revtex4-2}

\usepackage{graphicx}
\usepackage{subfigure}
\usepackage{bm}
\usepackage{latexsym,amsthm,amssymb}
\usepackage{booktabs,threeparttable}
\usepackage{makeidx}
\usepackage{amsmath}
\usepackage{amsfonts}
\usepackage{amssymb}
\usepackage{setspace}
\usepackage{slashed}

\usepackage{tikz}

\usepackage{hyperref}
\hypersetup{
    colorlinks=true,
    linkcolor=blue,
    filecolor=magenta,
    urlcolor=cyan,
}

\usepackage{xcolor}
\usepackage{xspace}

\let\originalleft\left
\let\originalright\right
\renewcommand{\left}{\mathopen{}\mathclose\bgroup\originalleft}
\renewcommand{\right}{\aftergroup\egroup\originalright}

\newcommand{\la}{\lambda_\alpha}

\newcommand{\fastjet}{F\protect\scalebox{0.8}{AST}J\protect\scalebox{0.8}{ET}\xspace}
\newcommand{\rivet}{R\protect\scalebox{0.8}{IVET}\xspace}

\newcommand{\fjcontrib}{\textsf{fjcontrib}}

\newcommand{\pt}{\ensuremath{p_{\text{T}}}}

\def\beq{\begin{equation}}  
\def\eeq{\end{equation}}
\def\({\left(}
\def\){\right)}
\def\[{\left[}
\def\]{\right]}

\begin{document}

\title{Identification  of $b$ jets using QCD-inspired observables
}

\author{Oleh~Fedkevych}
\email{oleh.fedkevych@ge.infn.it}
\affiliation{Dipartimento di Fisica, Universit\`a di Genova and INFN, Sezione di Genova, Via Dodecaneso 33, 16146, Italy}
\author{Charanjit  K. Khosa}
\email{charanjit.kaur@bristol.ac.uk}
\affiliation{H.H. Wills Physics Laboratory, University of Bristol, Tyndall Avenue, Bristol BS8 1TL, United Kingdom}
\author{Simone~Marzani}
\email{simone.marzani@ge.infn.it}
\affiliation{Dipartimento di Fisica, Universit\`a di Genova and INFN, Sezione di Genova, Via Dodecaneso 33, 16146, Italy}
\author{Federico Sforza}
\email{fsforza@cern.ch}
\affiliation{Dipartimento di Fisica, Universit\`a di Genova and INFN, Sezione di Genova, Via Dodecaneso 33, 16146, Italy}

\begin{abstract} 
We study the issue of separating hadronic jets that contain bottom quarks ($b$ jets) from jets featuring light partons only. 
We develop a novel approach to $b$ tagging that exploits the application of QCD-inspired jet substructure observables such as one-dimensional jet angularities and the two-dimensional primary Lund plane. 
We demonstrate that these observables can be used as inputs to modern machine-learning algorithms to efficiently separate $b$ jets from light ones. 
In order to test our tagging procedure, we consider simulated events where a $Z$ boson is produced in association with jets and show that using jet angularities as an input for a deep neural network, as well as using images obtained from the primary Lund jet plane as input to a convolutional neural network, one can achieve tagging accuracy comparable with the accuracy of conventional track-based taggers. 
We argue that the complementary usage of the track-based taggers together with the ones based upon QCD-inspired observables could  improve $b$-tagging accuracy. 
    
\end{abstract}
\maketitle

One of the most common final states resulting from high-energy particle collisions features collimated sprays of hadrons. These so-called hadronic jets can be seeded by particles with very different properties. For instance, jets can result from the fragmentation, and subsequent hadronization, of very energetic partons (quarks and gluons) or from the hadronic decays of (boosted) heavy particles, such as the Higgs boson, the electroweak ($W,Z$) bosons or the top quark.
Therefore, the identification of hadronic jets is a key aspect in particle physics and, consequently, a vibrant field of theoretical and experimental research on jet substructure has emerged and flourished in the last decade.
Theoretical advances in jet-substructure physics have led to the development of QCD-inspired jet observables that aim to efficiently distinguish signal from background, while maintaining desirable properties, such as resilience against difficult-to-model nonperturbative effects (see~\cite{Marzani:2019hun} and references therein).
Furthermore, in the past few years, the particle physics community has started actively using novel machine learning (ML) algorithms, which turned up to be very beneficial and, consequently, they are driving many of the most recent development in the fields; see \cite{Larkoski:2017jix, Feickert:2021ajf}.

In the context of jet tagging, an efficient and pure experimental selection of jets containing bottom quarks, henceforth $b$ jets, versus jets produced by other light flavors, dubbed $b$-jet tagging, is crucial for studies of the Higgs boson properties~\cite{ATLAS:2020FCP, CMS:2018nsn,ATLAS:2018mme,CMS:2018uxb}, measurements of Standard Model (SM) processes~\mbox{\cite{ATLAS:2020juj,ATLAS:2020aln,CMS:2018fks}}, and searches for beyond SM (BSM) phenomena~\mbox{\cite{LHCb:2021trn,ATLAS:2020xzu, CMS:2020pyk}}. 
The experimental identification of $b$ jets is possible exploiting the long lifetime ($\tau\approx 1.5$ ps), large mass ($m \gtrsim  5$ GeV), and decay patterns of the produced $b$ hadrons which are reconstructed thanks to  existing precision particle tracking detectors and complex multivariate algorithms~\cite{ATLAS:2019bwq, CMS:2017wtu}. 
However, because of the specific nature of the problem, very few studies have attempted to exploit theory-inspired observables to tackle the issue of $b$ tagging. 

In this paper, we suggest a novel approach to \mbox{$b$ tagging} using QCD-inspired observables and ML techniques. 
In particular, we use jet angularities~\cite{Berger:2003iw,Almeida:2008yp,Larkoski:2014pca} and the primary Lund plane (PLP)~\cite{Dreyer:2018nbf} as inputs to train ML algorithms. 
These observables have been the target of recent measurement campaigns at the LHC~\cite{ATLAS:2019mgf, ATLAS:2020bbn, CMS:2021iwu, ALICE:2021njq} as well as of detailed theoretical investigation~\cite{Caletti:2021oor,Reichelt:2021svh,Hornig:2016ahz,   Kang:2018qra, Kang:2018vgn,Lifson:2020gua,Dreyer:2021hhr}. 
Therefore, jet angularities and PLP provide us with the opportunity to construct taggers that are, at the same time, efficient and robust.
Because of the intrinsically multivariate nature of the $b$-jet tagging problem, we use the aforementioned  QCD-inspired observables to train a deep neural network (DNN) and a convolutional neural network (CNN), suited also for image recognition, in order to obtain optimal $b$-jet tagging performance from the combination of multiple features. 
The results are finally compared against $b$-jet tagging performance as provided in detail by the ATLAS collaboration~\cite{ATLAS:2019bwq}.

In the following, we study the case of high transverse momentum ($\pt \ge 500$~GeV)~\footnote{The coordinate system is right-handed with origin considered at the interaction point of two colliding proton beams revolving in a circular collider, the $x$ axis pointing toward the center of the ring, the $y$ axis pointing upward, and the $z$ axis along the collision direction. A projection over the $x,y$ plane is dubbed transverse and identified by a subscript $T$. Cylindrical coordinates $(r,\phi)$ are used in the transverse plane, $\phi$ being the azimuthal angle around the $z$ axis. The pseudorapidity variable  is defined in terms of the polar angle $\theta$ as $\eta = -\ln \tan(\theta/2)$. Angular separation is measured in units of $\Delta R \equiv \sqrt{ (\Delta \eta)^2 + (\Delta \phi)^2}$.} 
jets produced in proton-proton collisions at \mbox{$\sqrt{s}=13$ TeV collision energy,} a challenging benchmark for experiments because $b$-jet tagging performance degrades for jets of transverse momentum  beyond several hundred  \mbox{GeV~\cite{ATLAS:2019bwq,CMS:2017wtu}} which is a consequence of the worsening experimental resolution for high $\pt$ charged particle track 
reconstruction~\cite{Grupen:2008}. Furthermore,  fragmentation tracks become more abundant, and this dilutes the discrimination power, even in the presence of an ideal detector.
Moreover, the large amount of data collected at \mbox{$\sqrt{s} = 13$ TeV} collision energy$-$up to an integrated luminosity of $\int\mathcal{L}=$140 fb$^{-1}$ for the major LHC experiments$-$constitutes an invaluable source of physics information that can be exploited, for instance, by means of jet-substructure observables, as it has been already done using jet angularities~\cite{ATLAS:2019mgf,CMS:2021iwu,ALICE:2021njq} and PLP~\cite{ATLAS:2020bbn, ALICE:2021yet}. 
We also note that about 20 times larger integrated luminosity is foreseen to be collected in the future, requiring the optimal extrapolation of $b$-tagging techniques to the very high \pt~region, where BSM sensitivity is expected.

In this work, we study the behavior of the proposed $b$-tagging algorithm exploiting computer-generated pseudodata produced with the {PYTHIA v8.303}  Monte Carlo (MC)~\cite{Sjostrand:2014zea}.
To check the stability of our predictions with respect to change of a MC model we also consider the \mbox{HERWIG v7.2.1} code~\cite{Bellm:2015jjp, Bellm:2019zci}.
In both programs we are using their default settings and leading-order matrix elements.
The results based upon PYTHIA MC are shown in the main text, while the HERWIG ones are available in Appendix.

We simulate a $Z$ boson  production in association with hadronic jets, as in  the recent CMS measurements~\cite{CMS:2021iwu} with slightly adjusted leading jet cuts to be consistent with the ATLAS jet selection cuts from~\cite{ATLAS:2019bwq}, which we aim to compare our results to. 
More precisely, we require at least one anti-$k_t, R = 0.4$ jet~\cite{Cacciari:2008gp} with rapidity $|y_{\rm jet}| < 2.5$ and  \mbox{$p_{\rm T,\rm jet} \ge 500$ GeV}.

We define the flavor of each selected jet  as $b$, $c$,  and light-jets using hadron-level quantities, replicating the standard experimental procedures illustrated by the \mbox{ATLAS} analysis~\cite{ATLAS:2019bwq}.
More precisely, jets are labeled as $b$ jets if they are matched to at least one weakly decaying $b$ hadron having \mbox{$\pt \geq 5$ GeV} within a cone of size $\Delta R = 0.3$ around the jet axis;
if no \mbox{$b$ hadrons} are found, then the same selection criteria are used to search for matching \mbox{$c$ hadrons}. 
A jet matched to a \mbox{$c$ hadron} is labeled as  a \mbox{$c$-jet}. 
After assigning $b$ and $c$ jet labels we dub the remaining jets as light jets. 
Because it is a common approach~\cite{ATLAS:2019bwq, CMS:2017wtu} to quantify the quality of $b$-tagging algorithms comparing $b$-jet vs. light-jet discrimination, we discard jets with $c$ labels from our sample.
We generate 100 K samples for both \mbox{$b$ jets} and  light jets, using the MC event generators described above.
Finally, we build jet substructure observables  before and after application of the SoftDrop grooming algorithm~\cite{Larkoski:2014wba} with $\beta = 0$ and $z_{\rm cut} = 0.1$ parameters.
The application of the SoftDrop  algorithm allows us to reduce the sensitivity of the observables to soft radiation, which is difficult to model. 
Groomed observables are, therefore, more resilient than standard ones against  hadronization, underlying event, pileup and detector effects.
In the following, we use the label SoftDrop to indicate whether this grooming algorithm has been applied to the jets.
To avoid complications due to the possible \mbox{$p_{\rm T}$-bin} migration caused by SoftDrop we always refer to the value of the transverse momentum of a SoftDrop jet being measured \textit{before} grooming.

\begin{figure*}[ht!]
\subfigure[]{
\includegraphics[width=0.31\textwidth]{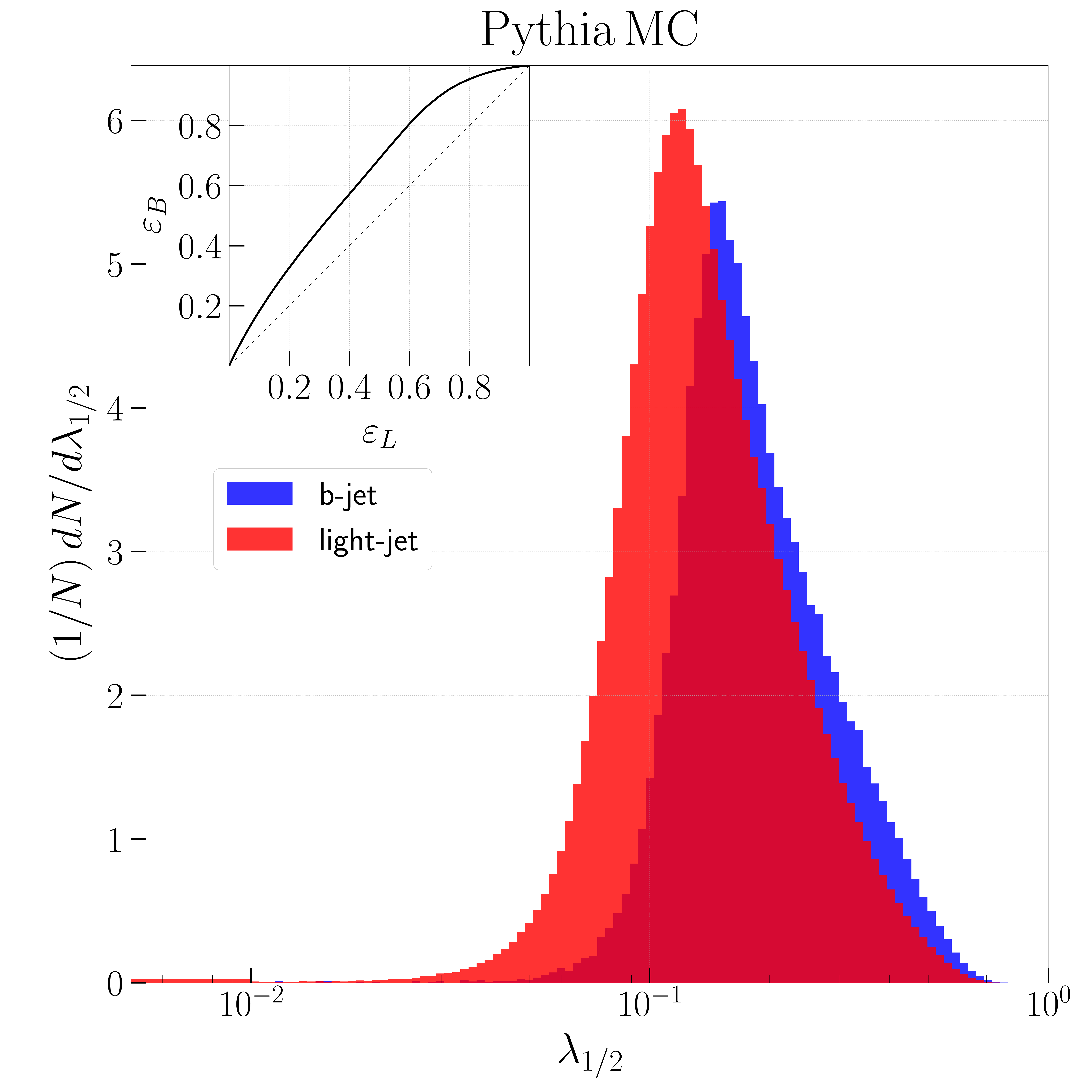}}
\subfigure[]{
\includegraphics[width=0.31\textwidth]{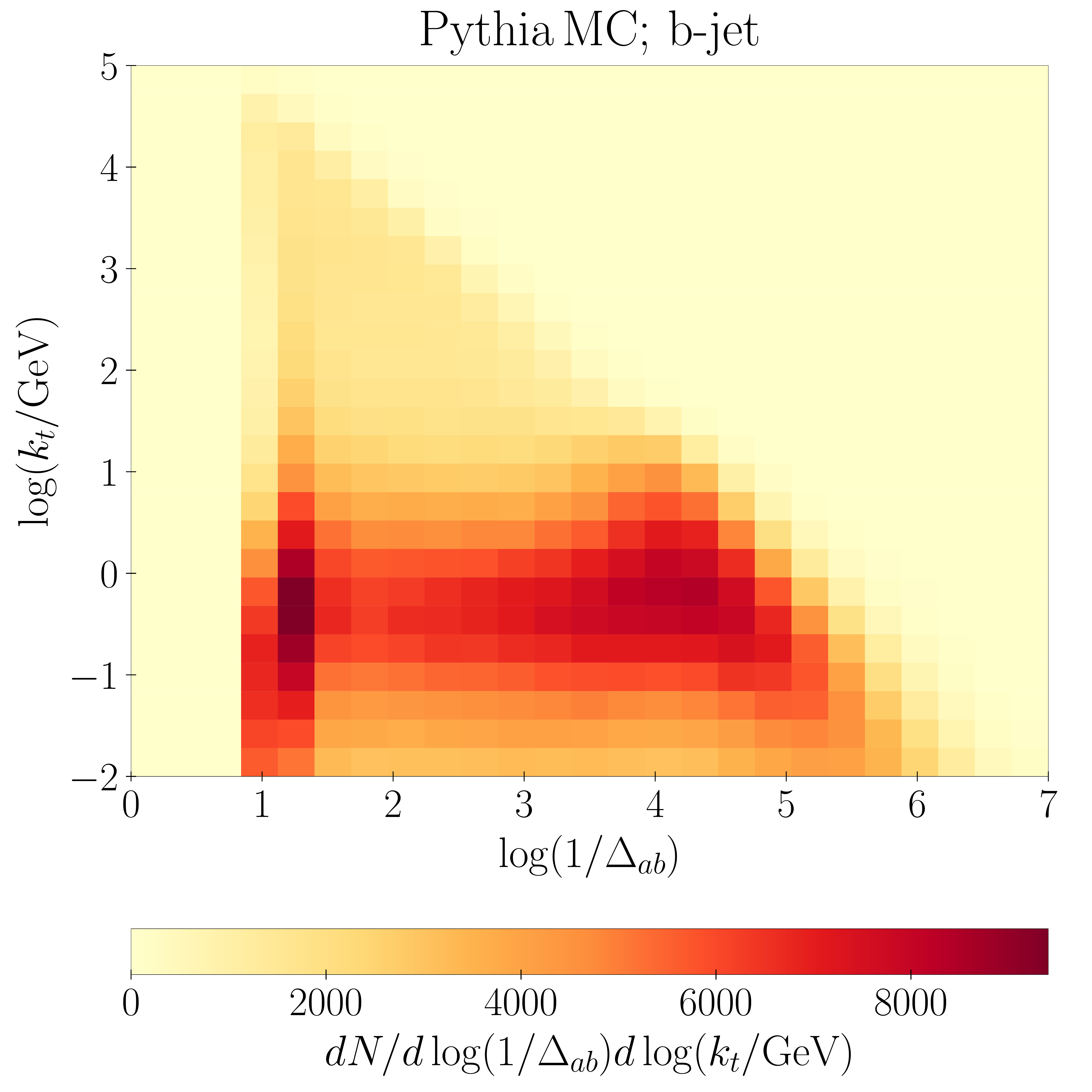}}
\subfigure[]{
\includegraphics[width=0.31\textwidth]{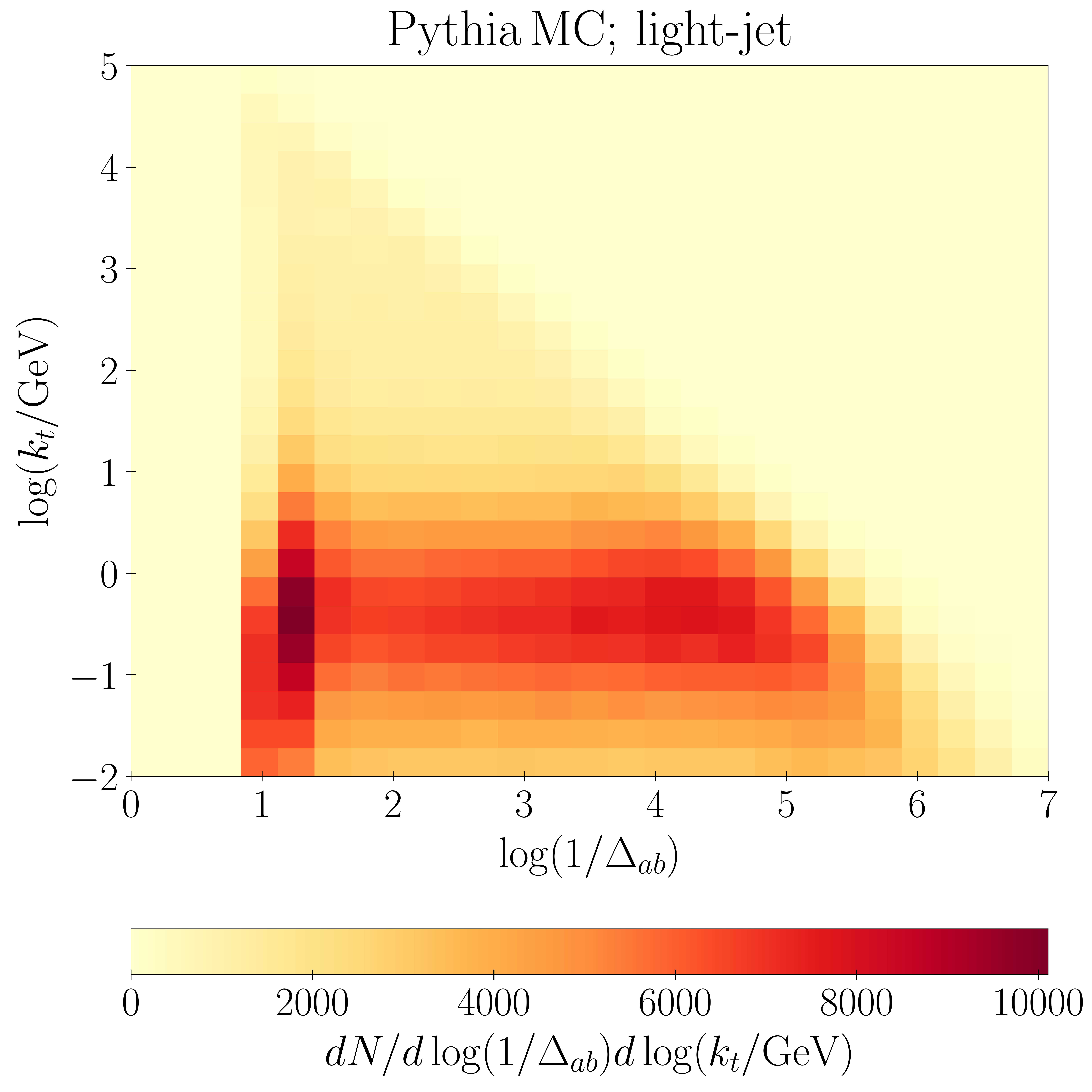}}\\
\subfigure[]{
\includegraphics[width=0.31\textwidth]{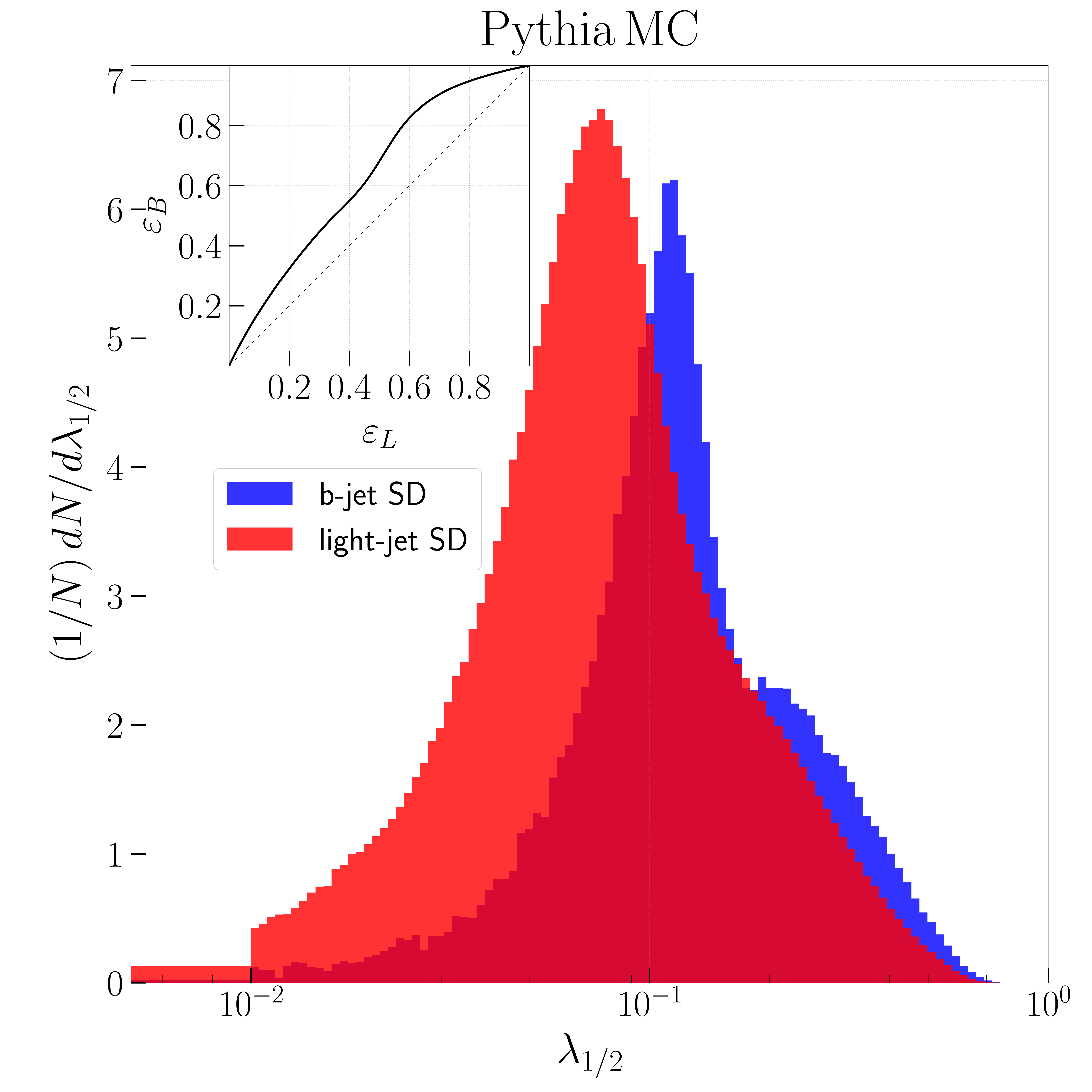}}
\subfigure[]{
\includegraphics[width=0.31\textwidth]{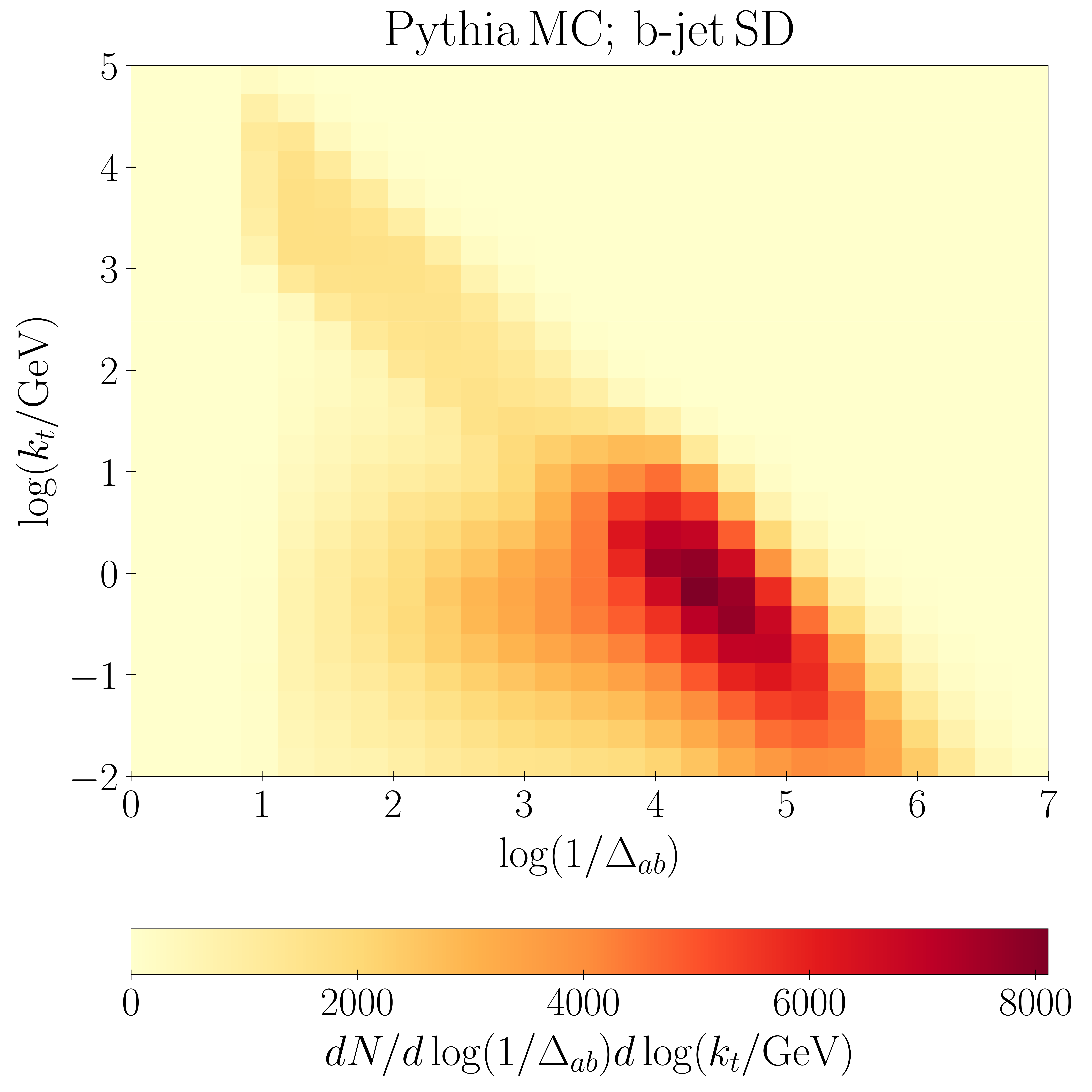}}
\subfigure[]{
\includegraphics[width=0.31\textwidth]{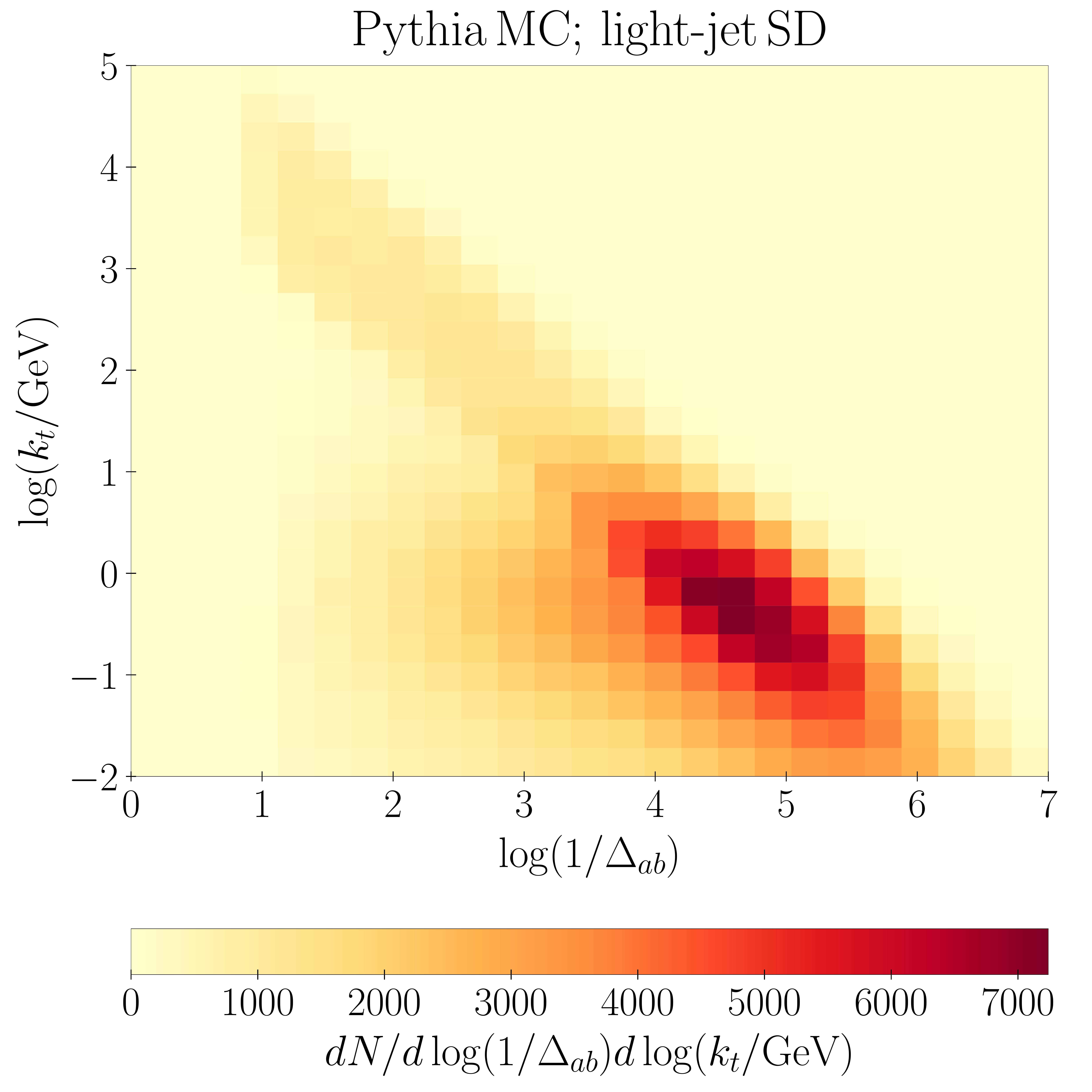}}
\caption{A collection of  jet substructure observables that we use in our analysis. (a) The Les Houches angularity, including the underflow events in the first bin; the PLP for $b$ jets (b) and light jets (c). The bottom row shows corresponding groomed distributions.}
  \label{fig:observables}
\end{figure*}

\begin{figure}
  \includegraphics[width=1.0\linewidth]{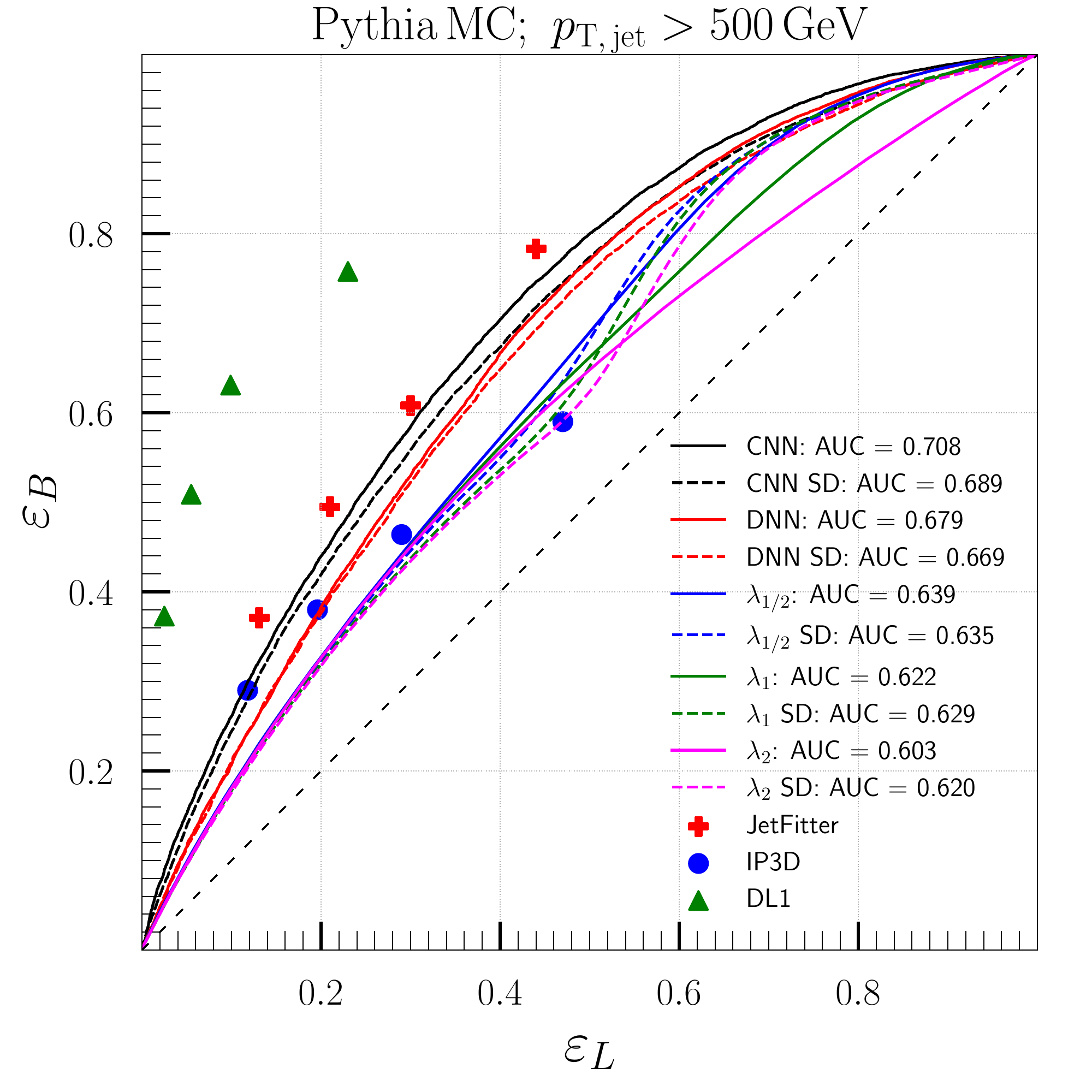}
  \caption{ROC curves for angularity distributions, multivariable DNN, and  PLP CNN classifiers compared against the ATLAS data~\cite{ATLAS:2019bwq}.}\label{fig:main_results}
\end{figure}

Let us first demonstrate how one can use the jet angularities to tell light jets from $b$ jets. 
Jet angularities are defined as 
\begin{equation}\label{eq:ang-def}
\la = \sum_{i \in \text{jet}}\left(\frac{p_{\rm T,i}}{\sum_{j \in \rm jet} p_{\rm T,j}}\right)\left(\frac{\Delta_i}{R} \right)^\alpha\,,
\end{equation}
where the sum runs over all jet constituents, $R$ is the jet radius, and 
\begin{equation}\label{eq:dist-def}
\Delta_i=\sqrt{(y_i-y_\text{jet})^2+(\phi_i-\phi_\text{jet})^2}\, 
\end{equation}
is the Euclidean azimuth-rapidity distance of particle $i$ from the jet axis.  
The requirement of infrared and collinear safety implies $\alpha > 0$. 
Therefore, we consider three commonly used cases namely, $\lambda_{1/2}$ (Les Houches angularity), $\lambda_1$ (jet width) and $\lambda_2$ (jet thrust). 
\mbox{In Figs.~\ref{fig:observables} (a) and (d)} we compare the Les Houches angularity  distribution for $b$- and light-jet samples produced with PYTHIA, before and after jet grooming (other distributions are available in Appendix).
The $b$-jet flavor discrimination performance is quantified using so-called receiver operating characteristic (ROC) curves,  shown in the inserts of Figs.~\ref{fig:observables} (a) and (d). 
The ROC curve is computed by varying a selection threshold along the observable, $\lambda_{\rm cut}$ in this case, and displaying in a two-dimensional plane the corresponding efficiency of selecting the $b$-jet signal, $\varepsilon_{B}$, versus the efficiency of selecting  the light-jet background,  $\varepsilon_{L}$, calculated according to
\begin{eqnarray}
	\varepsilon_{B/L} = 
	\frac{1}{N_{B/L}} \int\limits^1_{\lambda_{\rm cut}}
	\frac{dN_{B/L}}{d\lambda} d\lambda.
	\label{eq:roc_curve_def}
\end{eqnarray}
In addition to considering single pairs $\varepsilon_L$ - $\varepsilon_B$ to estimate the tagging performance of a given observable, one can also consider the area under the ROC curve (AUC) which gives an intuitive measure of the quality of a discriminating algorithm (it is maximal and equal to 1 for perfect discrimination between signal to background, i.e. $\varepsilon_B=1$ and $\varepsilon_L=0$  for any value of $\lambda_{\rm cut}$).
In Fig.~\ref{fig:main_results} we show the full set of ROC curves and AUC values for $\lambda_{1/2}$, $\lambda_1$, and $\lambda_2$ angularities calculated before and after grooming.
We see that the most efficient tagger based upon single jet angularity provides about 58\%  $b$-jet tagging efficiency versus about 40\% light-jet tagging efficiency, reaching an  area under the curve (AUC) of about 0.64. Similar efficiency is achieved after the application of the SoftDrop algorithm.

In order to improve over the simple cut-and-count approach using a single observable, we  
train two discriminants, one for jets before and after grooming correspondingly. 
Each DNN takes as an input values of three different jet angularities $\lambda_{1/2}$,  $\lambda_1$, and  $\lambda_{2}$. 
The DNN architecture  consists of two hidden layers having  five nodes in each layer.
More complex DNN architectures have also been tested without observing significant gains.
The ``ReLu'' and ``SoftMax'' activation functions are used for the intermediate and output layer, respectively.
To train the DNN we use Adam optimizer~\cite{Kingma:2014vow} and the cross-entropy loss function.
We used 60\% of the balanced data set for the training of the DNN, and the remaining 20\% for the validation and 20\% for the evaluation.
For each test event the DNN returns an output score in the interval $[0,1]$, with higher values if the event is more likely to be signal.
Therefore, it is possible to evaluate the ROC curve and the AUC of the DNN algorithm proceeding with a threshold scan over the DNN output scores. 

The results are shown in Fig.~\ref{fig:main_results}; we can see a relevant improvement in the performance as compared to the \mbox{cut-and-count} approach over any of the individual angularity distributions:
the ROC reaches a value of $\varepsilon_B\simeq 64$\% for a light-jet background efficiency of  $\varepsilon_L\simeq 40$\%, and the AUC reaches a value of 0.67, with no difference when considering the computation for jets before or after grooming.

The second QCD-inspired observable we have tested and found sensitive to the jet flavor is the PLP  originally introduced in Ref.~\cite{Andersson:1988gp} and recently applied to projection of a single jet in Ref.~\cite{Dreyer:2018nbf}. 
The PLP already found many applications, e.g., in the context of tagging~\cite{Dreyer:2018nbf, Dreyer:2020brq, Dreyer:2021hhr, Khosa:2021cyk, Khosa:2021zbx, Cavallini:2021vot},  dynamical grooming~\cite{Mehtar-Tani:2019rrk, Mehtar-Tani:2020oux}, generative models~\cite{Carrazza:2019cnt}, reinforced ML \cite{Carrazza:2019efs}, and  unsupervised new physics searches~\cite{Dillon:2020quc}. 
In the context of $b$ physics, Ref.~\cite{Cunqueiro:2018jbh} first proposed to use the PLP to look for the so-called \mbox{dead-cone} effect, i.e., the suppression of QCD radiation around massive quarks. This effect was then subsequently measured by the \mbox{ALICE} collaboration~\cite{ALICE:2021aqk}.
Furthermore, the all-order structure of the PLP density was recently computed in Ref.~\cite{Lifson:2020gua}.
Following Ref.~\cite{Dreyer:2018nbf}, we build the PLP by reclustering the selected jet using the Cambridge-Aachen  jet algorithm~\cite{Dokshitzer:1997in, Wobisch:1998wt}. 
We then follow the declustering history of the hardest branch and record at each splitting: 
\begin{eqnarray}
	k_t &\equiv& p_{\text{T}b} \, \Delta_{ab},
\end{eqnarray}
where $\Delta_{ab}$ is the azimuth-rapidity distance between subjets $a$ and $b$. 
In Figs.~\ref{fig:observables} (b), (c), (e), and (f) we show the two-dimensional PLP distributions (plotted on a log-log plane) for $b$ and light jets before  and after SoftDrop grooming.

Next, we explore the construction of a $b$-tagging algorithm based on PLP. Instead of a DNN, we now employ a CNN, which is better-suited for image datasets~\cite{VALUEVA2020232} .
As in the DNN training, we use 60\% of the $b$-jet signal and light-jet background events to train two CNN discriminants, both for groomed and ungroomed jets.
The CNN architecture consists of four convolutional layers followed by the flat layer. 
The First and second convolutional layers have 20 and 10 filters, respectively. The third and fourth layers have 8 filters. 
The flat layer has 200 neurons. 
The activation function and optimizers are the same as the ones used for the DNNs.

The ROC curves and corresponding AUC for the CNN algorithms are reported in Fig.~\ref{fig:main_results}, showing a relevant improvement over the DNN results.
The ROC reaches a value of $\varepsilon_B\simeq 70$\%, for a signal efficiency which corresponds to a light-jet background selection of  \mbox{$\varepsilon_L\simeq 40$\%}.  The AUC reaches a value of 0.71, with a decrease of 0.02 for the case of the groomed jets.

To check the robustness of our approach against the details of the simulation tools, e.g.\ the different approaches to model nonperturbative effects, we apply the CNN PLP tagger, trained upon PYTHIA inputs, to a dataset produced with the aforementioned HERWIG setup. 
The resulting ROC curves have profiles similar to those shown in Fig.~\ref{fig:main_results}. 
The difference between AUC values is varying in the 1$-$3\% interval (see also the Appendix).
These results give us confidence about the discrimination power of our proposed tagger. 

Finally, the performance of our DNN and CNN discriminants  is compared to the state-of-the-art $b$-tagging algorithms used by the ATLAS experiment.
The approach commonly followed by experimental collaboration is to combine a set of low-level $b$-tagging algorithms, based on detector reconstructed quantities, into high-level multivariate algorithms.
In the case of the \mbox{ATLAS} experiment the low-level $b$-tagging algorithms we could analyze are the JetFitter algorithm~\cite{ATLAS:2018nnq}, which attempts the reconstruction of the $b$- to $c$-hadron decay chain using fully or partially reconstructed vertices obtained from a subset of charged tacks associated to the jets, and  the  IP3D~\cite{ATLAS:2017bcq} algorithm, which analyzes, for each charged particle track, the three-dimensional distance of minimal approach between the proton-proton interaction vertex and the track trajectory. 
The information obtained by these algorithms (plus additional jet and secondary vertex reconstruction information) is used as an input for the so-called high-level $b$ tagger, in this case a deep feed-forward neural network named DL1~\citep{ATLAS:2017bcq}. 
Scatter points in Fig.~\ref{fig:main_results}  show the results of \mbox{$b$-jet} signal efficiency and light-jet background efficiency for the JetFitter, IP3D, and DL1 algorithms for jets reconstructed using PYTHIA-generated events, subsequently fed into the ATLAS Geant4-based detector simulation, plus charged particle track reconstruction.
The values are taken from~\cite{ATLAS:2010arf}.
By comparing the performance of the JetFitter, IP3D and DL1 taggers against our results we see that our DNN discriminants show better performance than the IP3D tagger and somewhat worse performance than the JetFitter tagger. 
However, we  see that our CNN discriminator improves our tagging performance and makes it comparable with the performance of the JetFitter algorithm. 
Finally, we note that the DL1 tagger, which is trained upon multiple features, leads to a  better performance than our DNN and CNN models.
Nevertheless, the set of simple input features considered in this article, which rely on QCD phenomenological ideas and are not directly based on charged particle track reconstruction or $b$-hadron decay properties,
can be used in conjunction with the aforementioned  ATLAS taggers, to improve existing multivariate tagging algorithms.
In order to support this statement we have checked that the output of the PLP with CNN tagger has no correlation with $b$-hadron decay distance from the proton collision or the invariant mass reconstructed using $b$-hadron decay charged particles.

We stress that the above discussion should be taken with some caution, as we are comparing information obtained with simulated data at the detector level by the \mbox{ATLAS} collaboration, with our own simulations, which are performed at particle level. 
However, we have a good level of confidence that this comparison is meaningful, because our taggers are built with infrared and collinear safe observables that should be robust against detector reconstruction inefficiency for
low momentum particles or single hadron reconstruction in dense environments.
For instance, related studies on Higgs tagging using the PLP with CNN~\cite{Cavallini:2021vot} have shown that detector effects typically result in a degradation of the tagging efficiency, as measured by the AUC, of a few percent, in the case of jets without grooming.
Furthermore, we expect  these effects to be even smaller, in the case of SoftDrop jets.

In this article, we have proposed a novel $b$-tagging approach based upon QCD-inspired jet substructure observables: one-dimensional jet angularities and a two-dimensional primary Lund plane. 
We have found that deep neural network and convolutional neural network discriminators trained upon these observables for jets of high transverse momentum  ($\pt \ge 500$~GeV) reach accuracy similar to $b$-tagging algorithms based on charged particle track reconstruction, as used by the ATLAS collaboration, but not as good as a more complex multivariate tagger which combines the aforementioned track-based tagger information.
Nevertheless, the advantage of our approach lies in its simplicity, since the only information one needs to use is the jet clustering history. 
Furthermore, both jet angularities and the primary Lund plane are sensitive to the kinematics of the jet constituents and their dynamics, as dictated by QCD. 
Both effects are influenced by the mass of the \mbox{$b$ quark}. 
On the other hand, existing \mbox{$b$ taggers} heavily rely on the \mbox{$b$-hadron} lifetime and decay properties, which are driven by electroweak physics. 
Therefore,  in the future, it would be especially interesting to combine the discriminating features discussed in this article with the ones already exploited in $b$-jet tagging algorithms used in experiments.

\section*{Acknowledgments}
We thank Andrea Coccaro for many useful comments on the manuscript.
The work of O.F. and S.M. is supported by Universit\`a di Genova under the curiosity-driven grant ``Using jets to challenge the Standard Model of particle physics'' and by the Italian Ministry of Research (MUR) under Grant No. PRIN 20172LNEEZ.
To build and train DNN and CNN networks we used Keras~\cite{chollet2015keras} and \mbox{TensorFlow~\cite{abadi2016tensorflow}} packages.
Our event selection and analysis are performed in the \rivet framework~\cite{Buckley:2010ar,Bierlich:2019rhm}, and we use \fastjet~\cite{Cacciari:2011ma} to cluster final-state hadrons into jets.
We also use the \fjcontrib~implementation of the SoftDrop groomer.
Figures  were created with the Matplotlib~\cite{Hunter:2007ouj} and NumPy~\cite{2020NumPy-Array} libraries.  
Our analysis files, NN models, and event samples are available upon request.

\clearpage

\appendix

\section*{Appendix}

In the following, several figures and plots offer additional details about the cross-checks and studies supporting our findings.

\begin{figure}[h!]
	\begin{minipage}[t]{0.45\linewidth}
        \centering
        \scalebox{0.6}{
\ifx\du\undefined
  \newlength{\du}
\fi
\setlength{\du}{15\unitlength}
\begin{tikzpicture}[even odd rule]
\pgftransformxscale{1.000000}
\pgftransformyscale{-1.000000}
\definecolor{dialinecolor}{rgb}{0.000000, 0.000000, 0.000000}
\pgfsetstrokecolor{dialinecolor}
\pgfsetstrokeopacity{1.000000}
\definecolor{diafillcolor}{rgb}{1.000000, 1.000000, 1.000000}
\pgfsetfillcolor{diafillcolor}
\pgfsetfillopacity{1.000000}
\definecolor{dialinecolor}{rgb}{0.000000, 0.000000, 0.000000}
\pgfsetstrokecolor{dialinecolor}
\pgfsetstrokeopacity{1.000000}
\definecolor{diafillcolor}{rgb}{0.000000, 0.000000, 0.000000}
\pgfsetfillcolor{diafillcolor}
\pgfsetfillopacity{1.000000}
\node[anchor=base west,inner sep=0pt,outer sep=0pt,color=dialinecolor] at (19.500000\du,17.000000\du){};
\pgfsetlinewidth{0.100000\du}
\pgfsetdash{}{0pt}
\pgfsetbuttcap
\pgfsetmiterjoin
\pgfsetlinewidth{0.100000\du}
\pgfsetbuttcap
\pgfsetmiterjoin
\pgfsetdash{}{0pt}
\definecolor{diafillcolor}{rgb}{1.000000, 0.000000, 0.000000}
\pgfsetfillcolor{diafillcolor}
\pgfsetfillopacity{0.392157}
\definecolor{dialinecolor}{rgb}{1.000000, 0.000000, 0.000000}
\pgfsetstrokecolor{dialinecolor}
\pgfsetstrokeopacity{1.000000}
\pgfpathmoveto{\pgfpoint{20.907824\du}{27.410899\du}}
\pgfpathlineto{\pgfpoint{27.175324\du}{27.410899\du}}
\pgfpathcurveto{\pgfpoint{28.040686\du}{27.410899\du}}{\pgfpoint{28.742199\du}{27.754458\du}}{\pgfpoint{28.742199\du}{28.178260\du}}
\pgfpathcurveto{\pgfpoint{28.742199\du}{28.602062\du}}{\pgfpoint{28.040686\du}{28.945621\du}}{\pgfpoint{27.175324\du}{28.945621\du}}
\pgfpathlineto{\pgfpoint{20.907824\du}{28.945621\du}}
\pgfpathcurveto{\pgfpoint{20.042463\du}{28.945621\du}}{\pgfpoint{19.340949\du}{28.602062\du}}{\pgfpoint{19.340949\du}{28.178260\du}}
\pgfpathcurveto{\pgfpoint{19.340949\du}{27.754458\du}}{\pgfpoint{20.042463\du}{27.410899\du}}{\pgfpoint{20.907824\du}{27.410899\du}}
\pgfpathclose
\pgfusepath{fill,stroke}
\definecolor{dialinecolor}{rgb}{0.000000, 0.000000, 0.000000}
\pgfsetstrokecolor{dialinecolor}
\pgfsetstrokeopacity{1.000000}
\definecolor{diafillcolor}{rgb}{0.000000, 0.000000, 0.000000}
\pgfsetfillcolor{diafillcolor}
\pgfsetfillopacity{1.000000}
\node[anchor=base,inner sep=0pt, outer sep=0pt,color=dialinecolor] at (24.041574\du,28.520565\du){$\bf Input \, 3$};
\pgfsetlinewidth{0.100000\du}
\pgfsetdash{}{0pt}
\pgfsetmiterjoin
\definecolor{diafillcolor}{rgb}{0.000000, 1.000000, 0.000000}
\pgfsetfillcolor{diafillcolor}
\pgfsetfillopacity{0.392157}
\fill (23.999960\du,43.000497\du)--(27.918119\du,47.979740\du)--(23.999960\du,52.958983\du)--(20.081800\du,47.979740\du)--cycle;
\definecolor{dialinecolor}{rgb}{0.000000, 1.000000, 0.000000}
\pgfsetstrokecolor{dialinecolor}
\pgfsetstrokeopacity{1.000000}
\draw (23.999960\du,43.000497\du)--(27.918119\du,47.979740\du)--(23.999960\du,52.958983\du)--(20.081800\du,47.979740\du)--cycle;
\definecolor{dialinecolor}{rgb}{0.000000, 0.000000, 0.000000}
\pgfsetstrokecolor{dialinecolor}
\pgfsetstrokeopacity{1.000000}
\definecolor{diafillcolor}{rgb}{0.000000, 0.000000, 0.000000}
\pgfsetfillcolor{diafillcolor}
\pgfsetfillopacity{1.000000}
\node[anchor=base,inner sep=0pt, outer sep=0pt,color=dialinecolor] at (23.999960\du,48.279254\du){$\bf Predictions$};
\pgfsetlinewidth{0.100000\du}
\pgfsetdash{}{0pt}
\pgfsetmiterjoin
\definecolor{diafillcolor}{rgb}{1.000000, 0.733333, 0.000000}
\pgfsetfillcolor{diafillcolor}
\pgfsetfillopacity{0.392157}
\pgfpathellipse{\pgfpoint{24.000000\du}{33.499980\du}}{\pgfpoint{5.054206\du}{0\du}}{\pgfpoint{0\du}{1.516262\du}}
\pgfusepath{fill}
\definecolor{dialinecolor}{rgb}{1.000000, 0.733333, 0.000000}
\pgfsetstrokecolor{dialinecolor}
\pgfsetstrokeopacity{1.000000}
\pgfpathellipse{\pgfpoint{24.000000\du}{33.499980\du}}{\pgfpoint{5.054206\du}{0\du}}{\pgfpoint{0\du}{1.516262\du}}
\pgfusepath{stroke}
\definecolor{dialinecolor}{rgb}{0.000000, 0.000000, 0.000000}
\pgfsetstrokecolor{dialinecolor}
\pgfsetstrokeopacity{1.000000}
\definecolor{diafillcolor}{rgb}{0.000000, 0.000000, 0.000000}
\pgfsetfillcolor{diafillcolor}
\pgfsetfillopacity{1.000000}
\node[anchor=base,inner sep=0pt, outer sep=0pt,color=dialinecolor] at (24.000000\du,33.799494\du){$\bf Dense$ $\bm N_1 = 5$};
\pgfsetlinewidth{0.100000\du}
\pgfsetdash{}{0pt}
\pgfsetbuttcap
{
\definecolor{diafillcolor}{rgb}{0.000000, 0.000000, 0.000000}
\pgfsetfillcolor{diafillcolor}
\pgfsetfillopacity{1.000000}
\pgfsetarrowsend{stealth}
\definecolor{dialinecolor}{rgb}{0.000000, 0.000000, 0.000000}
\pgfsetstrokecolor{dialinecolor}
\pgfsetstrokeopacity{1.000000}
\draw (24.035188\du,28.995812\du)--(24.012231\du,31.934386\du);
}
\pgfsetlinewidth{0.100000\du}
\pgfsetdash{}{0pt}
\pgfsetmiterjoin
\definecolor{diafillcolor}{rgb}{1.000000, 0.733333, 0.000000}
\pgfsetfillcolor{diafillcolor}
\pgfsetfillopacity{0.392157}
\pgfpathellipse{\pgfpoint{24.054206\du}{38.546591\du}}{\pgfpoint{5.054206\du}{0\du}}{\pgfpoint{0\du}{1.516262\du}}
\pgfusepath{fill}
\definecolor{dialinecolor}{rgb}{1.000000, 0.733333, 0.000000}
\pgfsetstrokecolor{dialinecolor}
\pgfsetstrokeopacity{1.000000}
\pgfpathellipse{\pgfpoint{24.054206\du}{38.546591\du}}{\pgfpoint{5.054206\du}{0\du}}{\pgfpoint{0\du}{1.516262\du}}
\pgfusepath{stroke}
\definecolor{dialinecolor}{rgb}{0.000000, 0.000000, 0.000000}
\pgfsetstrokecolor{dialinecolor}
\pgfsetstrokeopacity{1.000000}
\definecolor{diafillcolor}{rgb}{0.000000, 0.000000, 0.000000}
\pgfsetfillcolor{diafillcolor}
\pgfsetfillopacity{1.000000}
\node[anchor=base,inner sep=0pt, outer sep=0pt,color=dialinecolor] at (24.054206\du,38.846105\du){$\bf Dense$ $\bm N_2 = 5$};
\pgfsetlinewidth{0.100000\du}
\pgfsetdash{}{0pt}
\pgfsetbuttcap
{
\definecolor{diafillcolor}{rgb}{0.000000, 0.000000, 0.000000}
\pgfsetfillcolor{diafillcolor}
\pgfsetfillopacity{1.000000}
\pgfsetarrowsend{stealth}
\definecolor{dialinecolor}{rgb}{0.000000, 0.000000, 0.000000}
\pgfsetstrokecolor{dialinecolor}
\pgfsetstrokeopacity{1.000000}
\draw (24.016781\du,35.062261\du)--(24.037425\du,36.984310\du);
}
\pgfsetlinewidth{0.100000\du}
\pgfsetdash{}{0pt}
\pgfsetbuttcap
{
\definecolor{diafillcolor}{rgb}{0.000000, 0.000000, 0.000000}
\pgfsetfillcolor{diafillcolor}
\pgfsetfillopacity{1.000000}
\pgfsetarrowsend{stealth}
\definecolor{dialinecolor}{rgb}{0.000000, 0.000000, 0.000000}
\pgfsetstrokecolor{dialinecolor}
\pgfsetstrokeopacity{1.000000}
\draw (24.045197\du,40.113217\du)--(24.028738\du,42.975289\du);
}
\end{tikzpicture}}\\ a)
    \end{minipage}
    \begin{minipage}[t]{0.45\linewidth}
        \centering
        \scalebox{0.6}{\input{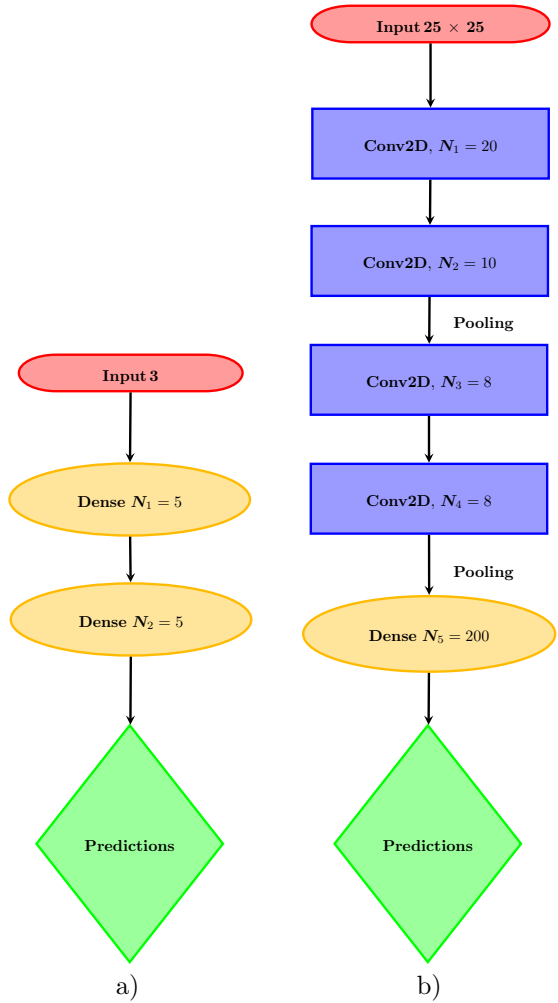}}\\ b)
    \end{minipage}
  \caption{Network architectures  we use. a)  DNN network.  b)  CNN network. Filter size is $3 \times 3$ in all convolutional layers.}
  \label{fig:nn_architecture}
\end{figure}

In Fig.~\ref{fig:nn_architecture} we provide two diagrams showing the details of the DNN/CNN architectures we use.

\begin{figure*}
\subfigure[]{
	\includegraphics[width=0.25\textwidth]{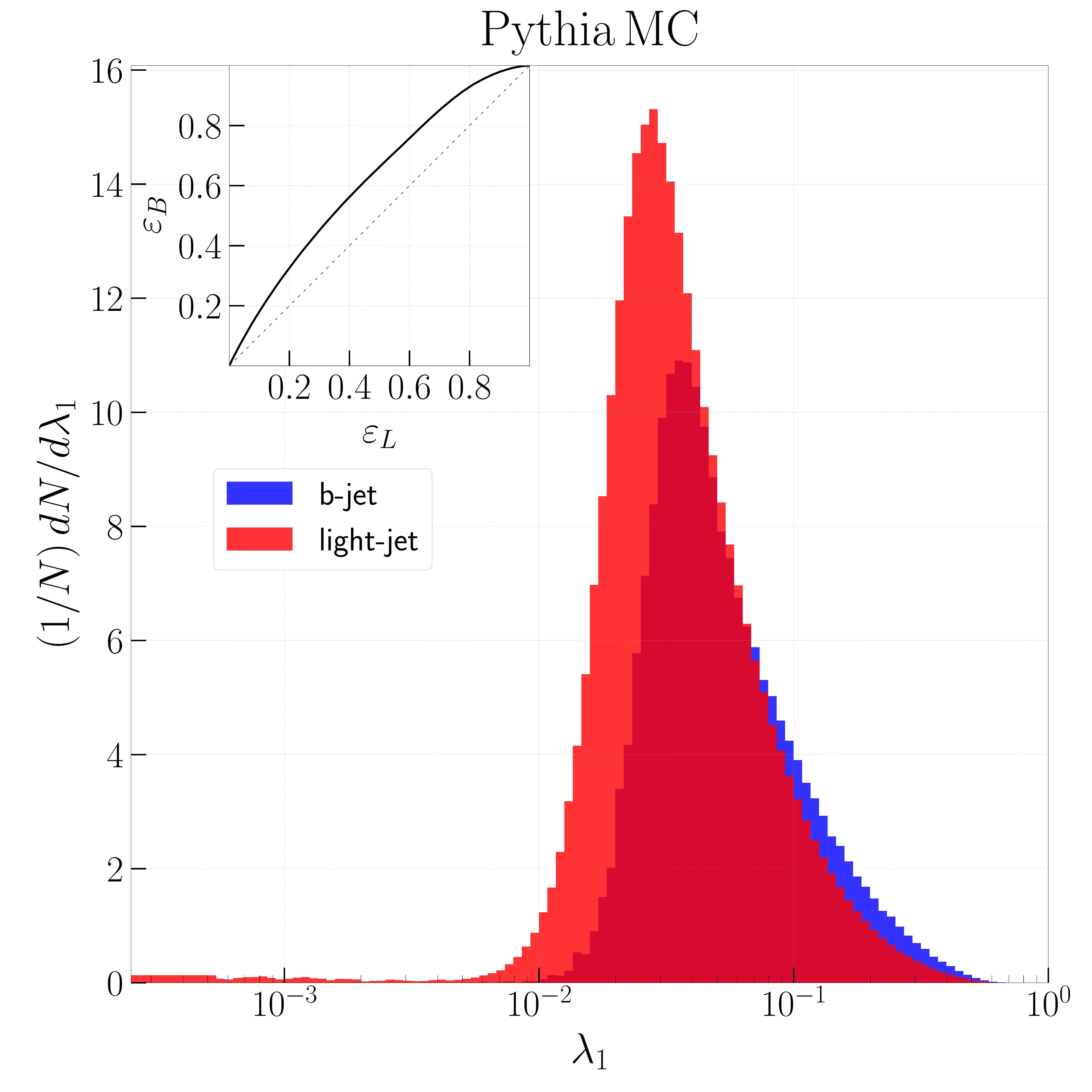}}
\subfigure[]{
	\includegraphics[width=0.25\textwidth]{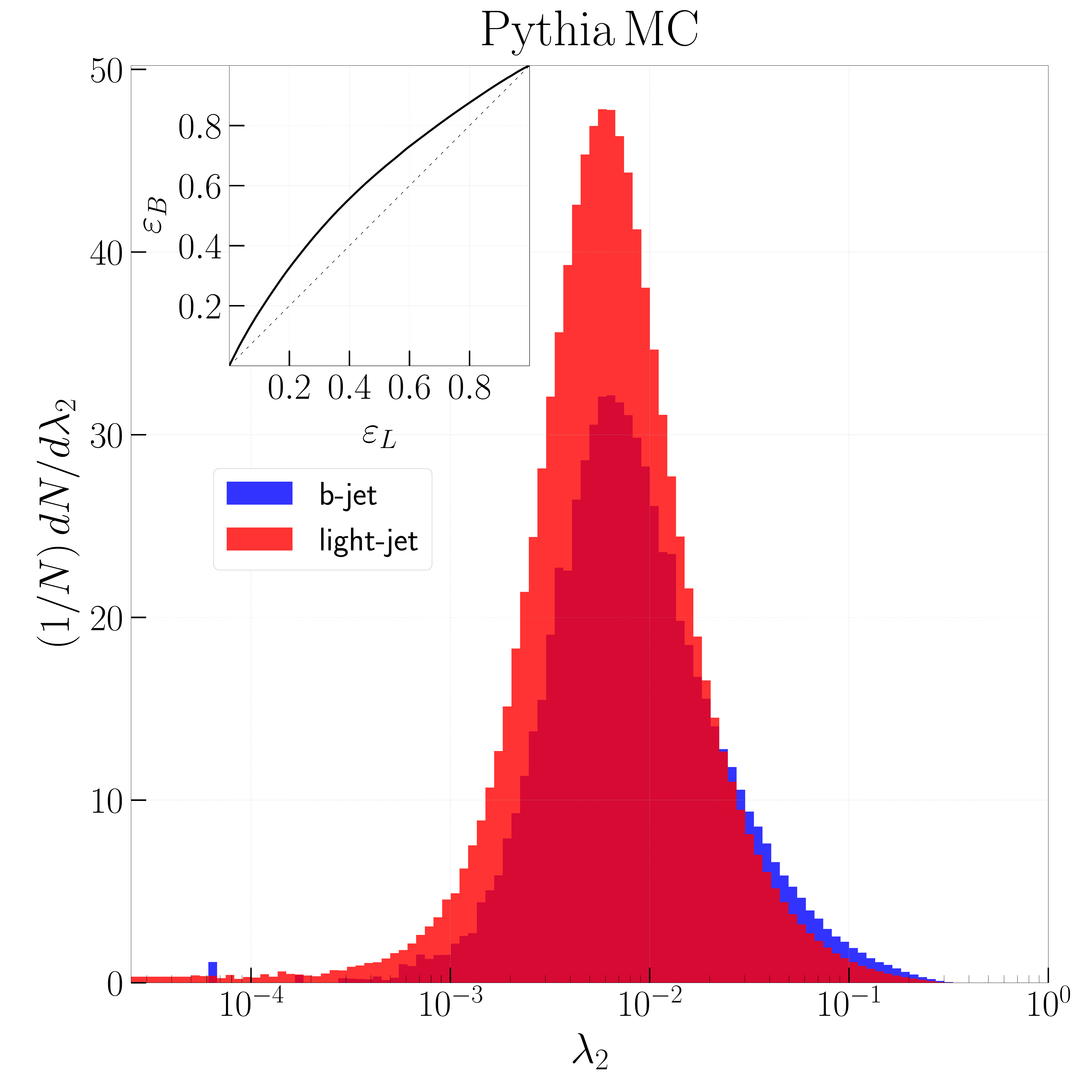}}\\
\subfigure[]{
ls	\includegraphics[width=0.25\textwidth]{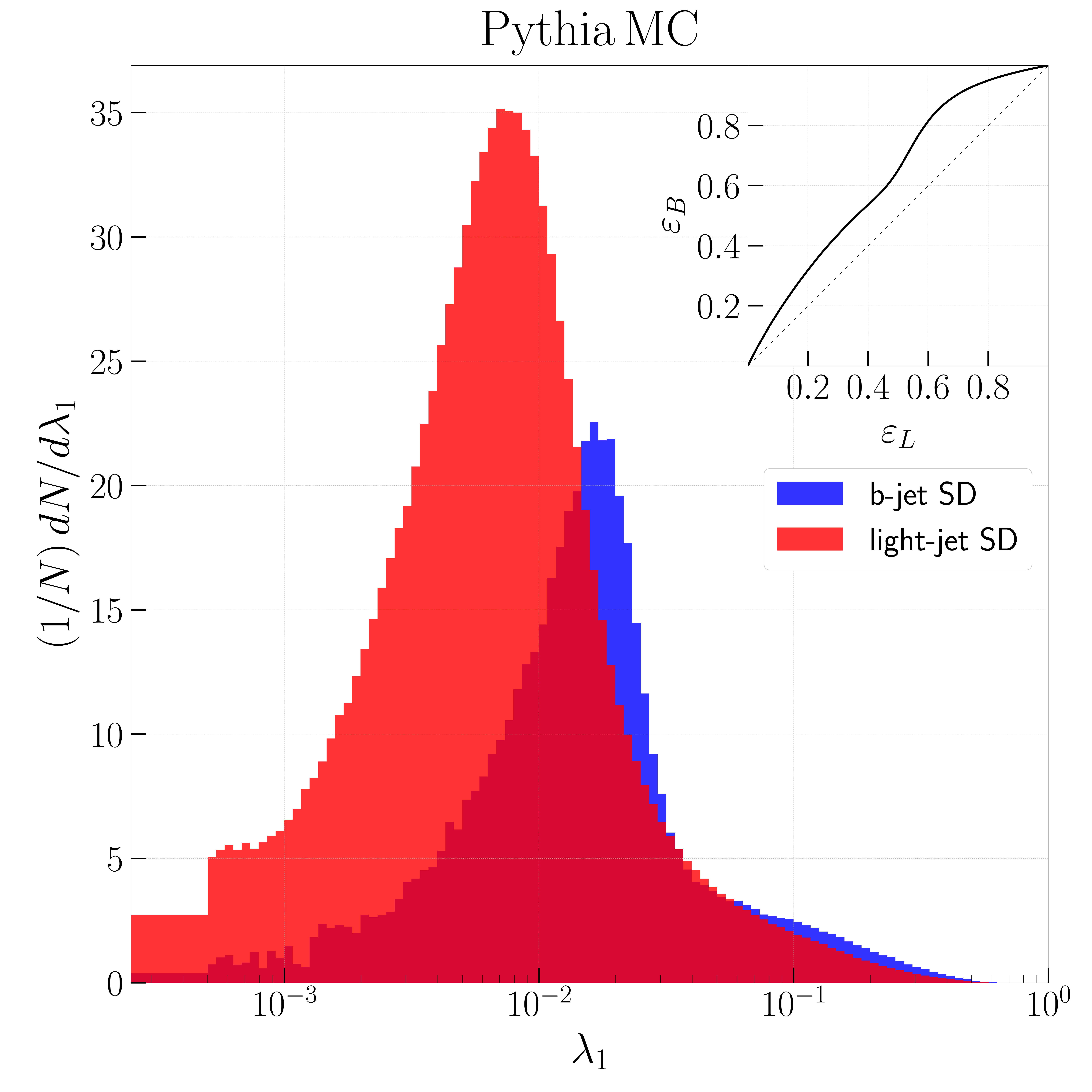}}
\subfigure[]{
	\includegraphics[width=0.25\textwidth]{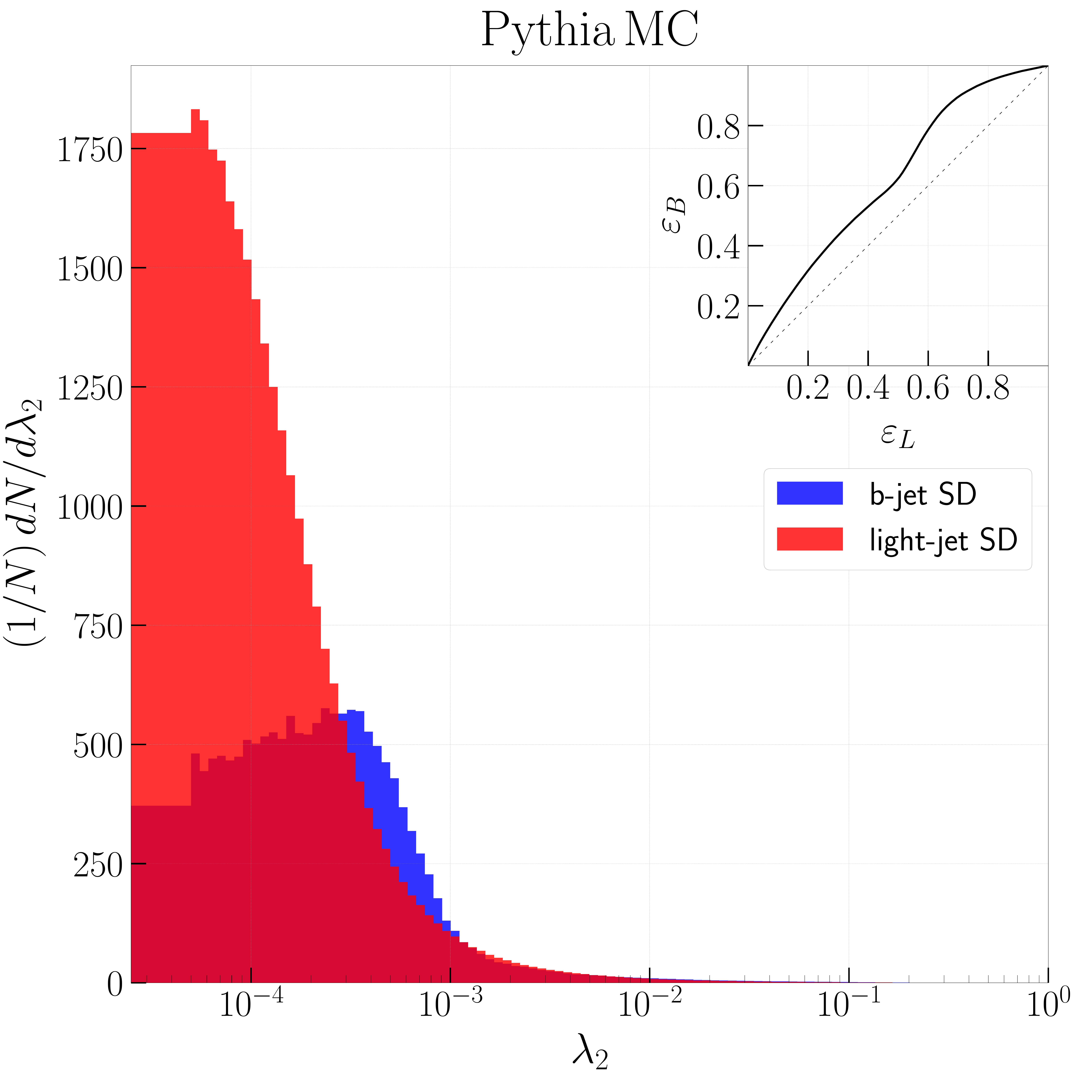}} \\
	\caption{Jet width ($\lambda_1$) and jet thrust ($\lambda_2$) distributions generated with the PYTHIA event generator. Upper row corresponds to ungroomed distributions whereas the lower row to groomed ones.}
	\label{fig:pythia_width_thrust}
\subfigure[]{
\includegraphics[width=0.25\textwidth]{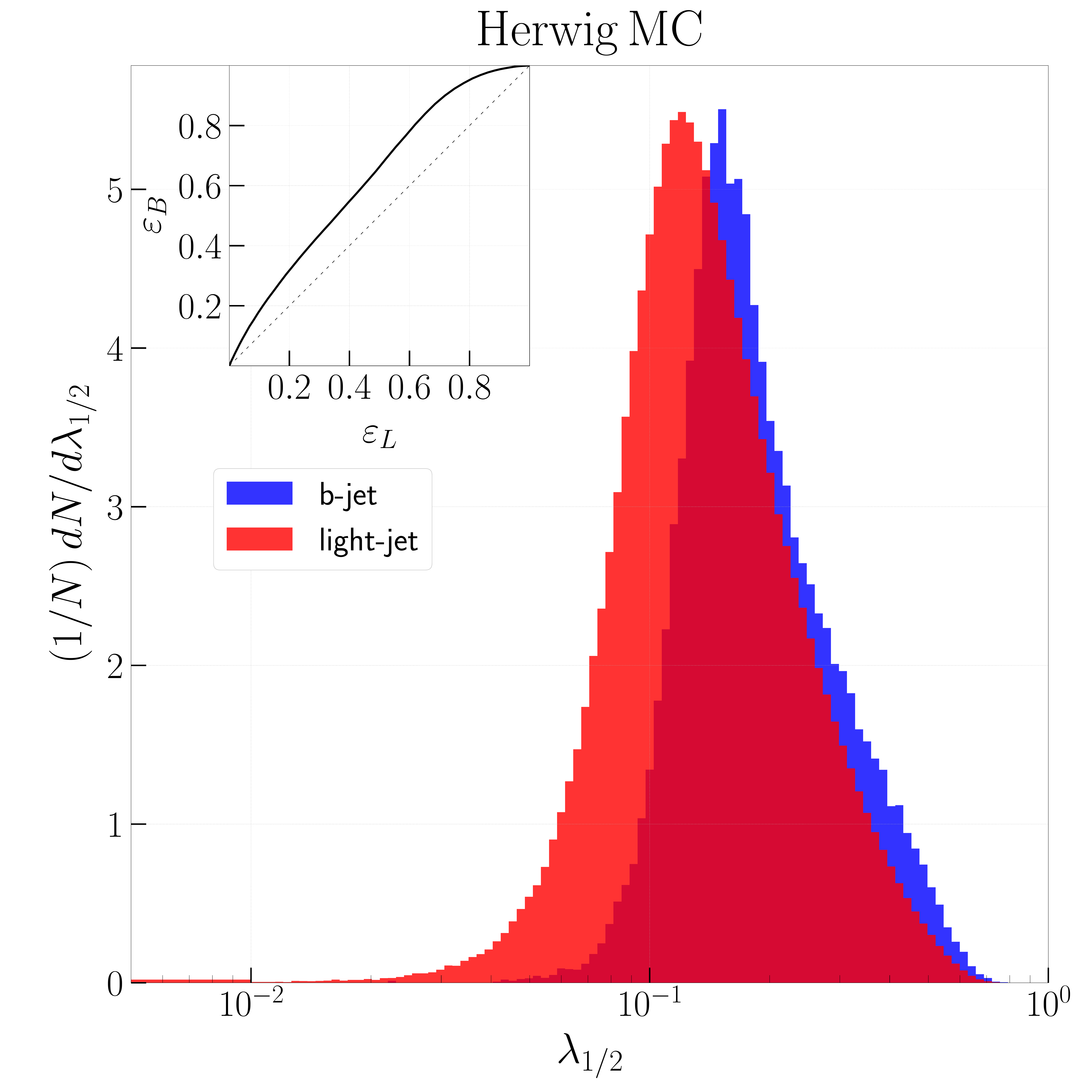}}
\subfigure[]{
\includegraphics[width=0.25\textwidth]{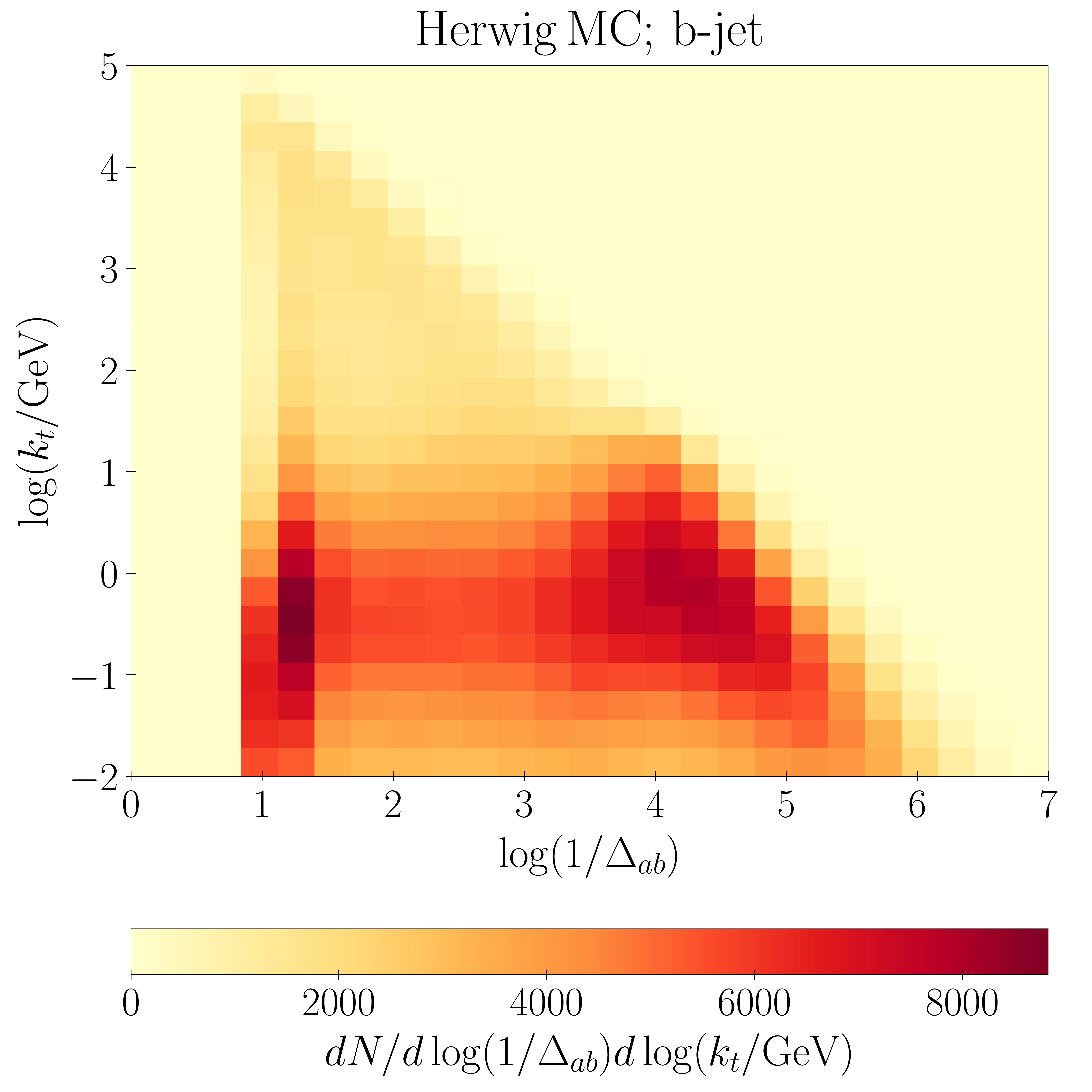}}
\subfigure[]{
\includegraphics[width=0.25\textwidth]{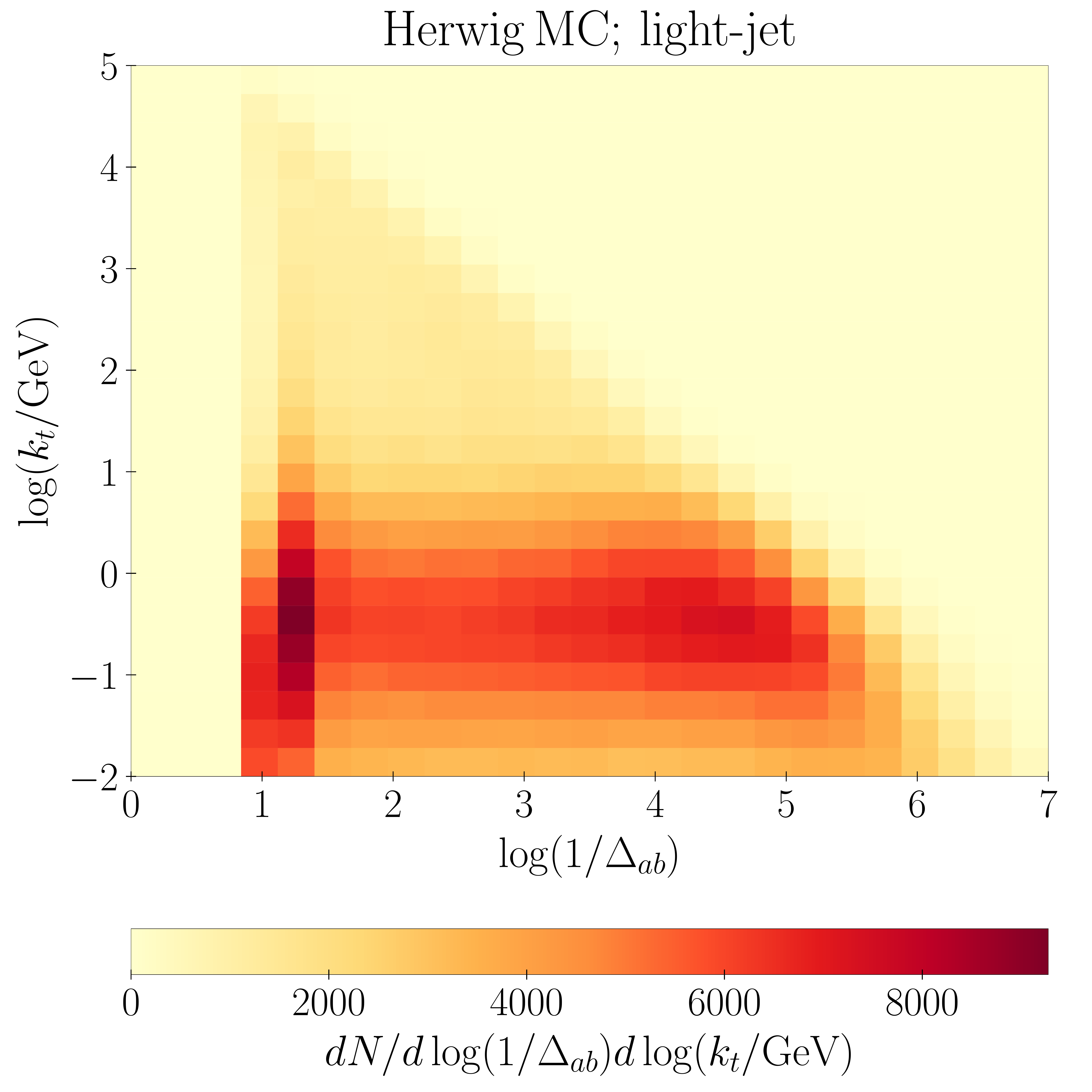}}\\
\subfigure[]{
\includegraphics[width=0.25\textwidth]{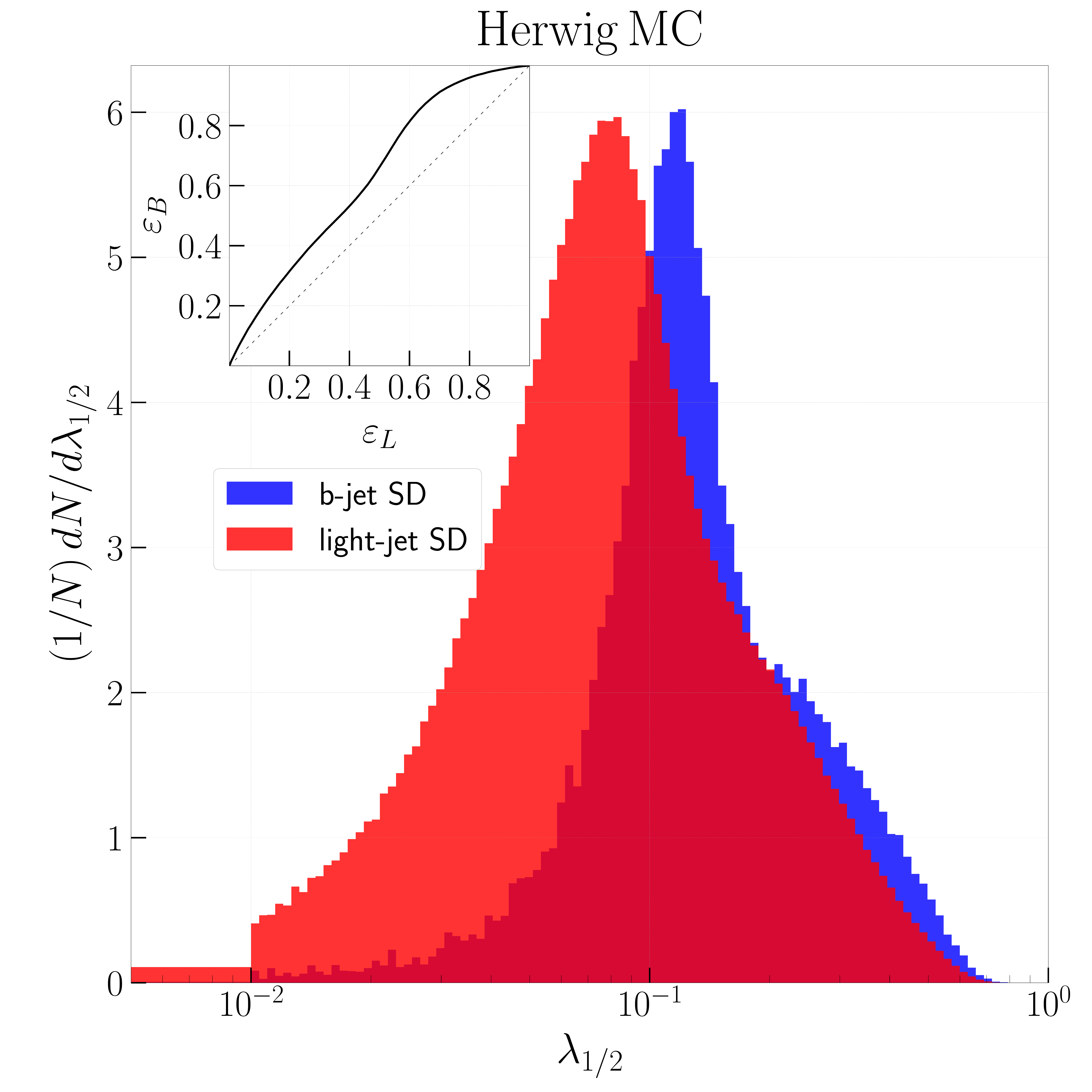}}
\subfigure[]{
\includegraphics[width=0.25\textwidth]{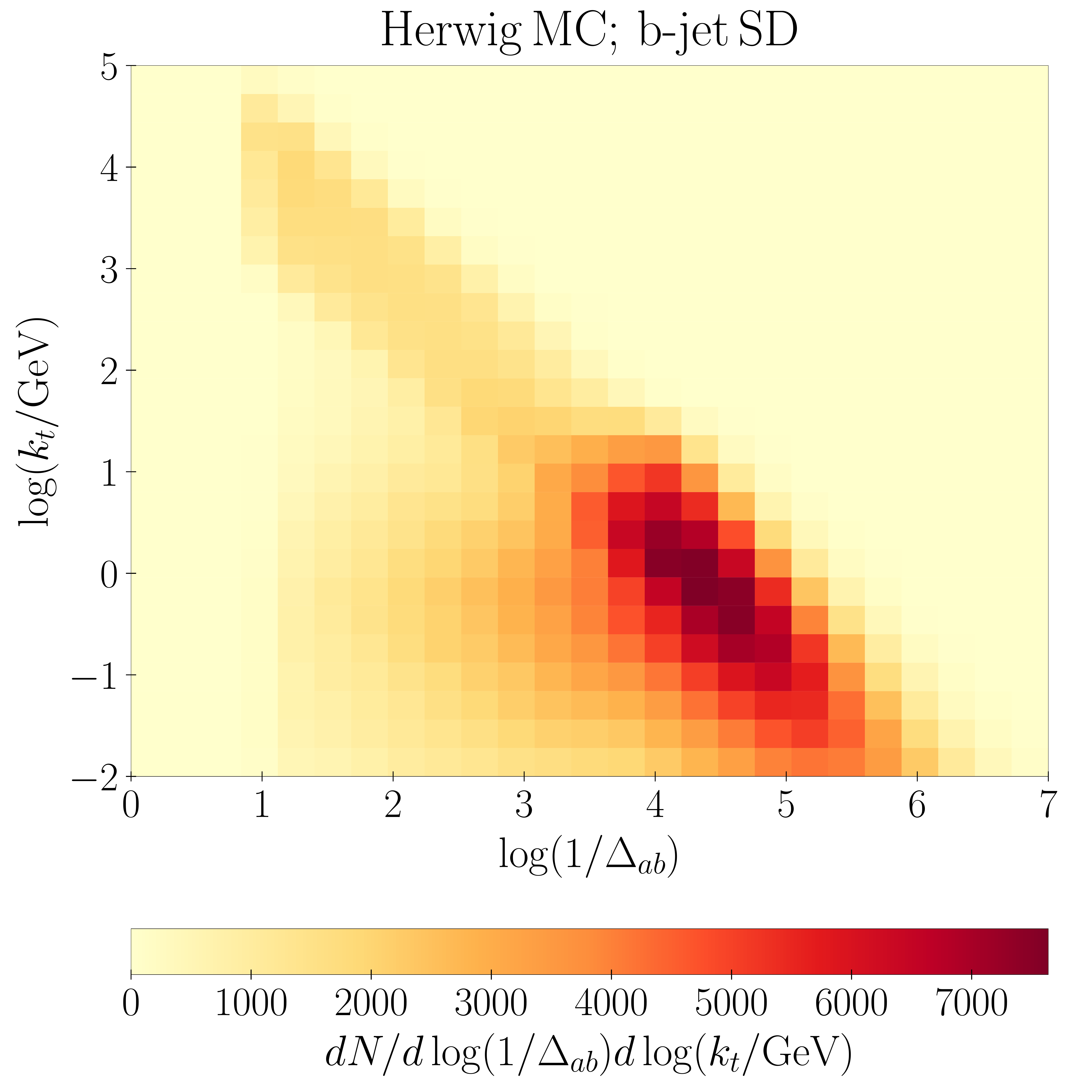}}
\subfigure[]{
\includegraphics[width=0.25\textwidth]{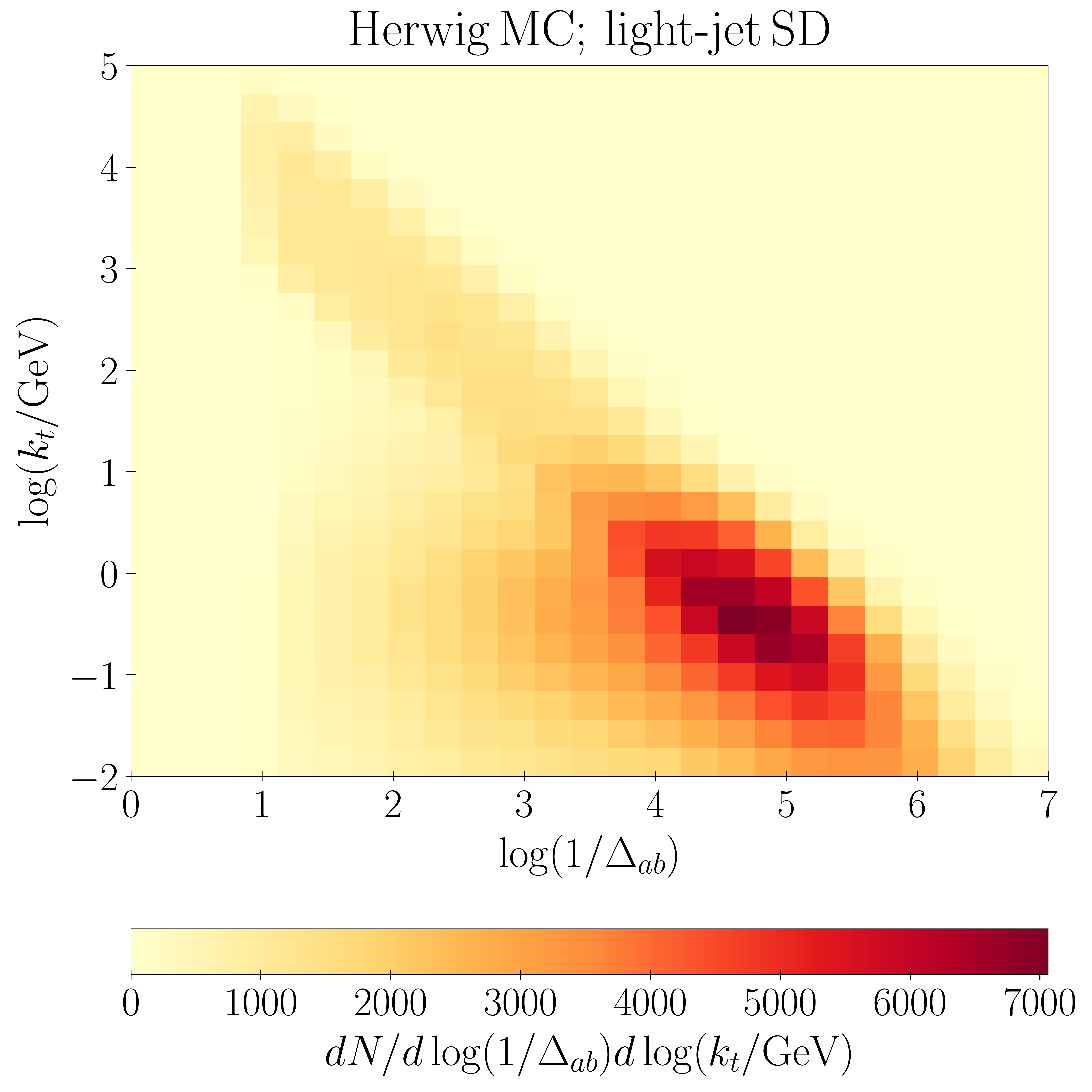}}
\caption{An example of different jet substructure observables we use for $b$ jets and light jets simulated with HERWIG MC.
    The top row shows the following for hadron-level jets: (a) the histogram of the Les Houches angularity  $\lambda_{1/2}$, including the underflow events in the first bin, and the PLP for $b$ jets (b) and light jets (c).
    The bottom row shows the following for hadron-level jets after the application of the SoftDrop grooming algorithm  with $\beta = 0$ and $z_{\rm cut} = 0.1$ parameters:
    (d) the histogram of the Les Houches angularity  $\lambda_{1/2}$, including the underflow events in the first bin, and the PLP for $b$ jets (e) and light jets (f).
  }
  \label{fig:herwig_lha_lund}
\end{figure*}

In Fig.~\ref{fig:pythia_width_thrust} we provide $\lambda_1$ and $\lambda_2$ distributions. 
The inserts in Fig.~\ref{fig:pythia_width_thrust} show the ROC curves for the one-dimensional taggers defined upon a  simple angularity cut  according to
\begin{eqnarray}
	\varepsilon_{B/L} = 
	\frac{1}{N_{B/L}} \int\limits^1_{\lambda_{\rm cut}}
	\frac{dN_{B/L}}{d\lambda} d\lambda.
	\label{eq:roc_curve_def}
\end{eqnarray}

To ensure that our findings are not sensitive to the details of simulation, all results obtained with PYTHIA have been cross-checked using the \mbox{HERWIG} MC program.
In Fig.~\ref{fig:herwig_lha_lund} we provide $\lambda_{1/2}$ and PLP distributions for $b$ jets and light jets before and after grooming obtained with the HERWIG MC.

\begin{figure}
  \includegraphics[width=1.0\linewidth]{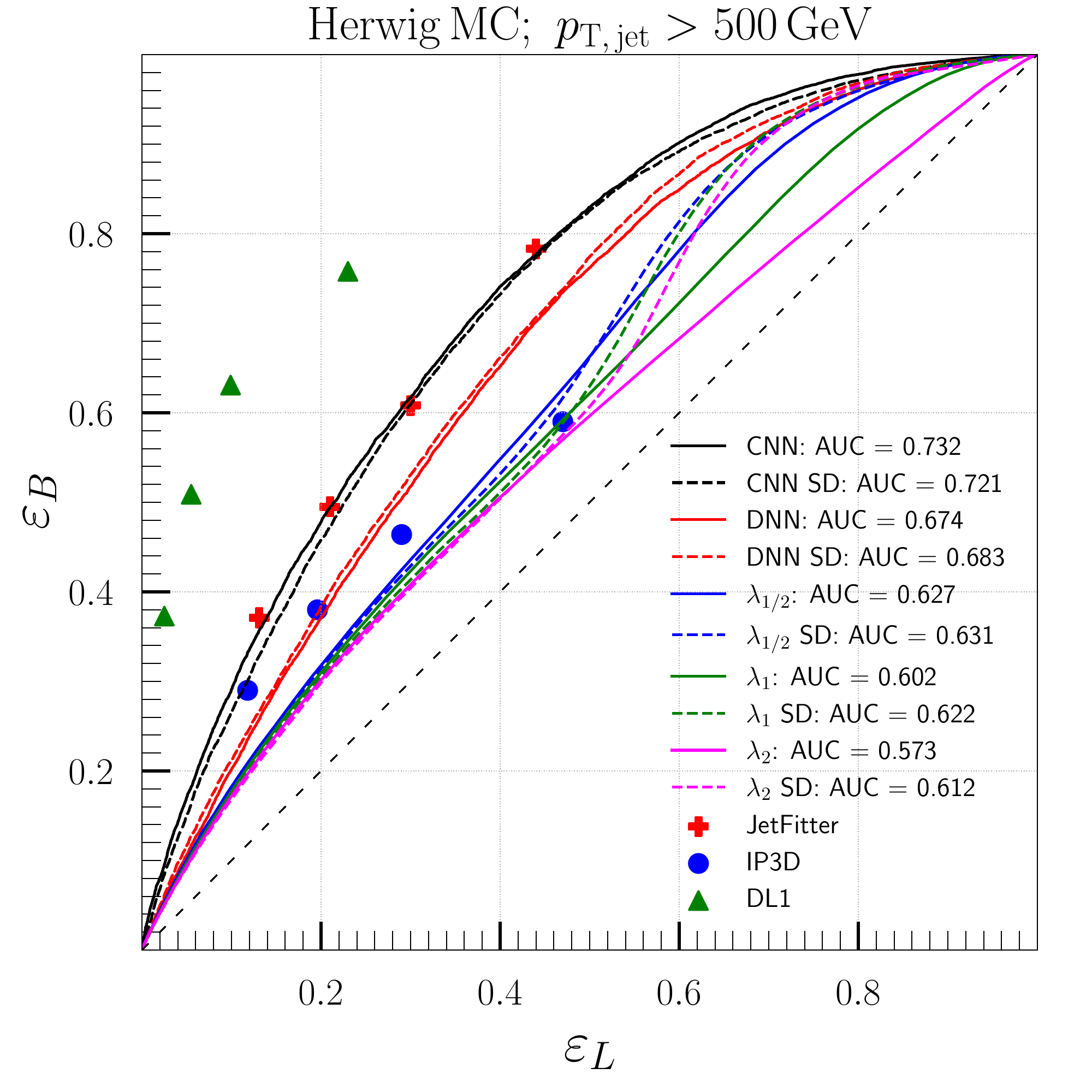}
   \caption{ROC curves, obtained using  HERWIG simulated data, for one-dimensional angularity distributions, multivariable DNN classifier, and  PLP CNN classifier. The single points correspond to ATLAS SV1, IP3D, and DL1 $b$-tagging performance.}
  \label{fig:main_results_check}
\end{figure}

\begin{figure}
  \includegraphics[width=1.0\linewidth]{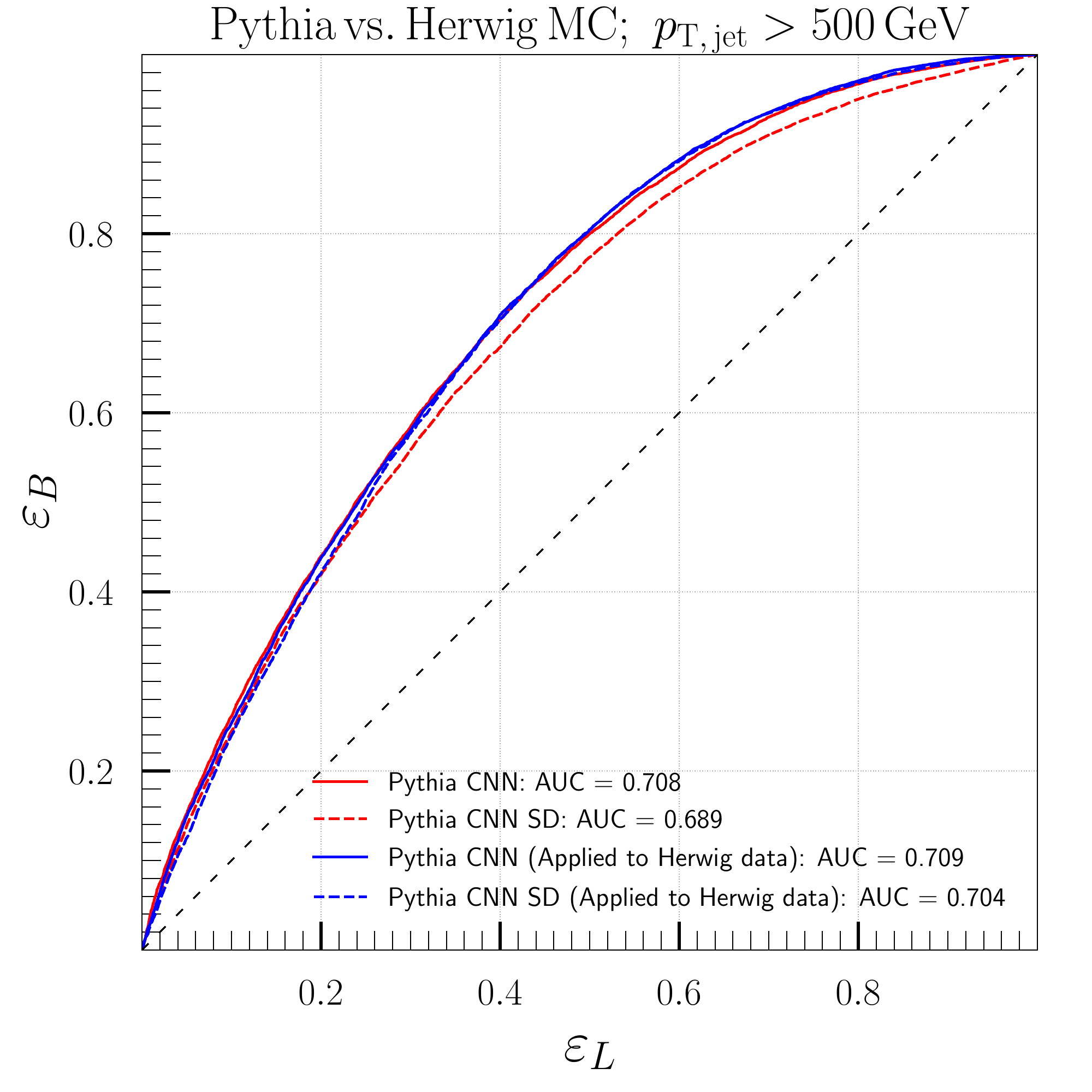}
   \caption{Robustness test of our CNN model. The ROC curves are obtained by applying the CNN trained upon PYTHIA data to the HERWIG data.}
  \label{fig:main_results_check_pythia_vs_herwig}
\end{figure}

We compare the performance of different taggers using HERWIG pseudodata in Fig.~\ref{fig:main_results_check}.
We see that the AUC values corresponding to the simple taggers based upon a single jet angularity observable differ by 2$-$5\% from the corresponding values obtained with the PYTHIA simulation, and this difference reduces to 1\% when considering SoftDrop jets.
We also note that the  differences between AUC values  obtained with DNN and CNN trained upon PYTHIA and HERWIG  are within 1$-$5\% interval.
We also shall note that  we observe somewhat different profiles of the ROC curves generated with different MC tools.
In Fig.~\ref{fig:main_results_check_pythia_vs_herwig} we provide the ROC curves obtained by applying the CNN trained up PYTHIA MC to the HERWIG data.
By calculating the corresponding AUC values and comparing them against the AUC values presented in the main part of our paper, we see that our predictions remain stable under the change of the MC dataset.
More precisely, we see that the AUC value for ungroomed  PLP data changes from  0.708 (the CNN trained and tested on the PYTHIA data) to 0.709 (the CNN trained on the PYTHIA MC and tested on the HERWIG data).
In the case of the groomed PLP distributions the AUC values change from 0.689 to 0.704, correspondingly. 

\begin{figure}[ht!]
	\includegraphics[width=0.49\textwidth]{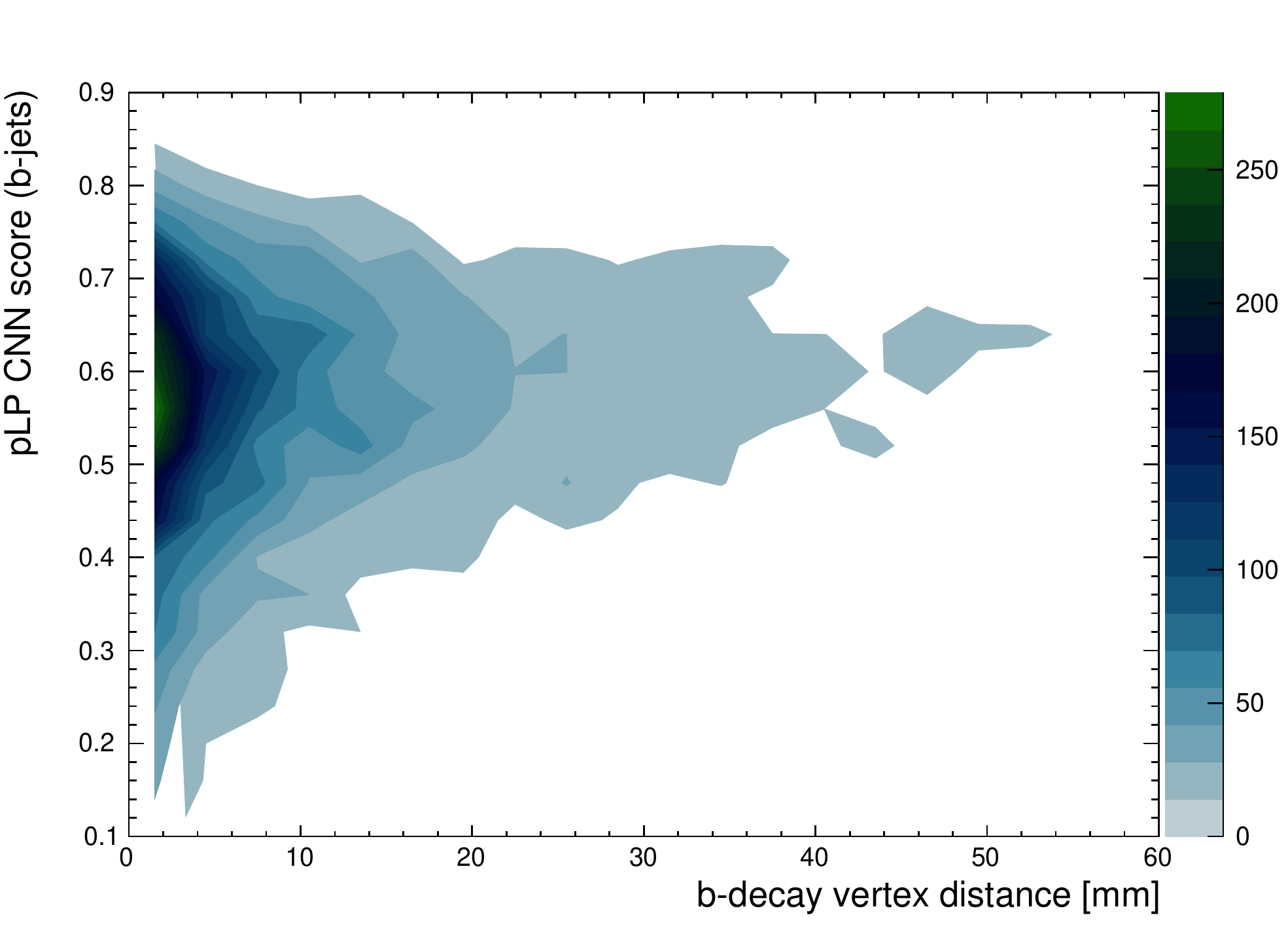}
		\includegraphics[width=0.49\textwidth]{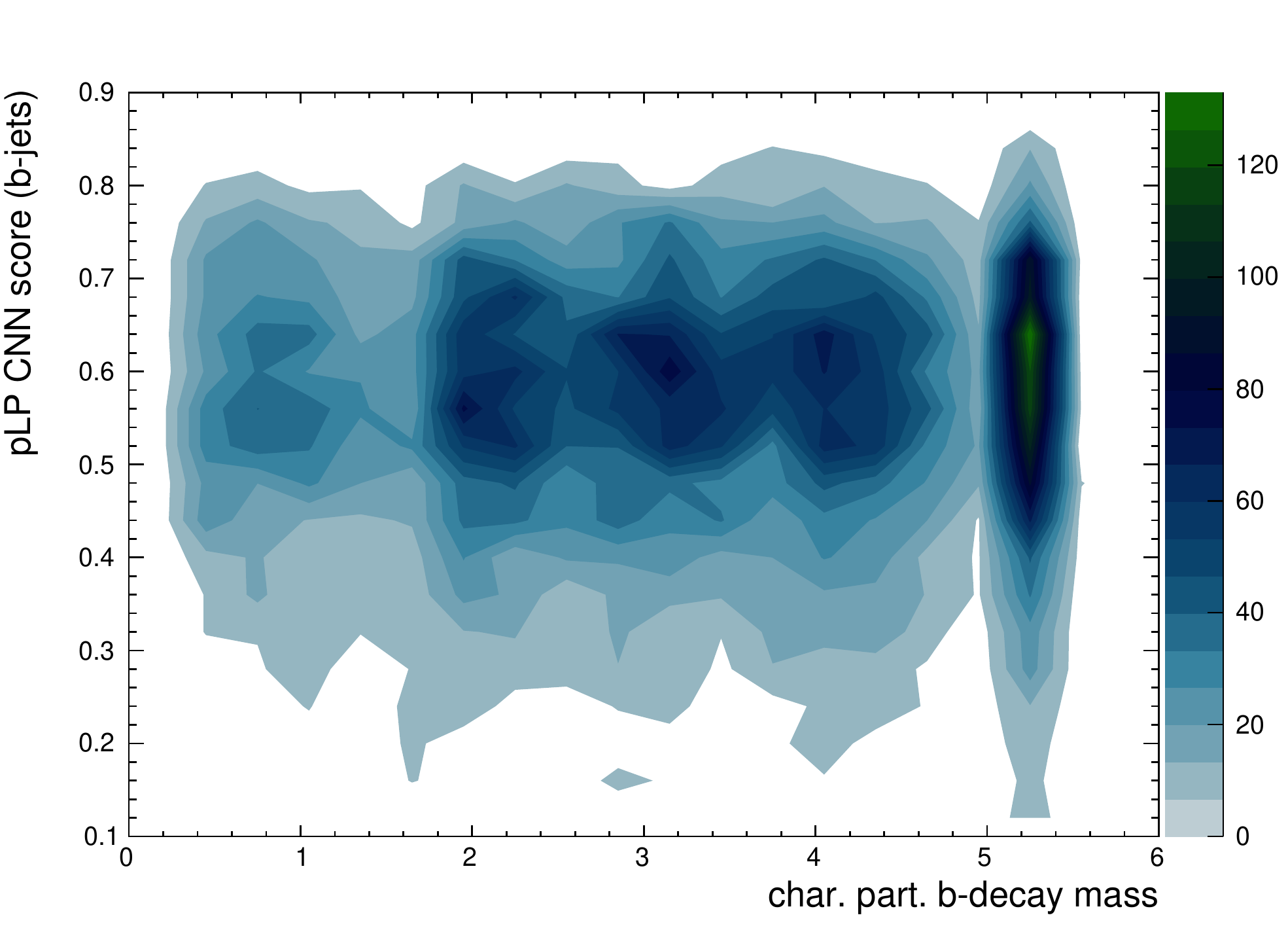}
	        \caption{PYTHIA 8 MC simulation of 10 K $Z+b$ jets  events with $b$ jet $p_T>500$~GeV. Both figures show, on the Y axis,
                  the CNN output score built using primary Lund plane information versus  the decay distance of the leading $b$ hadron on the (upper plot) or versus the invariant mass reconstructed using charged hadrons and leptons of $p_T>0.5$ GeV (a typical selection threshold in experiments)  originating from the $b$-hadron decay (lower plot).}
	\label{fig:correlation}
\end{figure}

Finally, Fig.~\ref{fig:correlation} summarizes the analysis of correlation between the CNN output score and key properties of the $b$-hadron decay as used by experiments.
Using \mbox{10 K} events of the \mbox{PYTHIA 8 MC} simulation (as used in the paper), Fig.~\ref{fig:correlation} shows the CNN output score distribution for $b$ jets on the Y axis versus the 
decay distance of the leading $b$ hadron on the X axis of the left plot, and versus the invariant mass reconstructed using charged hadrons and leptons of \mbox{$p_T>$ 0.5 GeV}
(a typical selection threshold in experiments)  originating from the $b$-hadron decay on the X axis of the right plot.
We found that the correlation coefficient between the tested distribution is  below 0.06 in all cases.

\clearpage

\bibliographystyle{apsrev4-2}
\bibliography{references}

\end{document}